\documentclass[12pt,a4paper,twoside]{article}
\usepackage{geometry}
\usepackage[round]{natbib}
\usepackage{cancel}
\usepackage{mathtools}
\allowdisplaybreaks
\usepackage{mathrsfs}
\usepackage{amsfonts}
\usepackage{amssymb}
\usepackage{booktabs}
\usepackage{graphicx}
\usepackage{placeins}
\usepackage{stmaryrd}
\usepackage{lipsum}  
\usepackage{epstopdf}
\usepackage{xcolor}
\usepackage{tikz}
\usepackage{setspace} 
\usepackage{bm}
\usepackage[labelfont=bf]{caption}
\usepackage[skip=-0.25\textwidth]{subcaption}
\usepackage{enumitem}
\usepackage{titlesec}

\newcommand{\magenmarker}{\protect\tikz[baseline=-0.5ex]\protect\draw[black, fill=magenta, line width=0.4pt] (0,0) circle (0.6ex);}
\newcommand{\bluemarker}{\protect\tikz[baseline=-0.5ex]\protect\draw[black, fill=blue, line width=0.4pt] (0,0) circle (0.6ex);}
\newcommand{\whitemarker}{\protect\tikz[baseline=-0.5ex]\protect\draw[black, fill=white, line width=0.4pt] (0,0) circle (0.6ex);}

\titleformat{\section}
{\normalfont\fontsize{12}{15}\bfseries}{\thesection .}{0.3em}{}
\titleformat{\subsection}
{\normalfont\fontsize{12}{15}\itshape\bfseries\centering}{\upshape\thesubsection .}{0.3em}{}
\titleformat{\subsubsection}
{\normalfont\fontsize{12}{15}\itshape\bfseries}{\upshape\thesubsubsection .}{0.3em}{}

\usepackage{color}
\usepackage[titletoc,title]{appendix}
\usepackage{xr-hyper}
\usepackage{hyperref}%
\hypersetup{
	colorlinks   = true,
	citecolor    = blue,
    linkcolor    = blue
}
\usepackage{xpatch}
\usepackage[normalem]{ulem}

\makeatletter
\patchcmd{\NAT@citex}
  {\@citea\NAT@hyper@{%
     \NAT@nmfmt{\NAT@nm}%
     \hyper@natlinkbreak{\NAT@aysep\NAT@spacechar}{\@citeb\@extra@b@citeb}%
     \NAT@date}}
  {\@citea\NAT@nmfmt{\NAT@nm}%
     \NAT@aysep\NAT@spacechar
     \NAT@hyper@{\NAT@date}}
  {}{}

\patchcmd{\NAT@citex}
  {\@citea\NAT@hyper@{%
     \NAT@nmfmt{\NAT@nm}%
     \hyper@natlinkbreak{\NAT@spacechar\NAT@@open\if*#1*\else#1\NAT@spacechar\fi}%
     {\@citeb\@extra@b@citeb}%
     \NAT@date}}
  {\@citea\NAT@nmfmt{\NAT@nm}%
     \NAT@spacechar\NAT@@open\if*#1*\else#1\NAT@spacechar\fi
     \NAT@hyper@{\NAT@date}}
  {}{}
\makeatother
\usepackage{placeins}
\renewenvironment{abstract}  
{\normalfont
	\begin{center}
		\bfseries \abstractname\vspace{-.5em}\vspace{0pt}
	\end{center}
	\list{}{%
		\setlength{\leftmargin}{0.5 mm}
		\setlength{\rightmargin}{\leftmargin}%
	}%
	\item\relax}
{\endlist}  
\def\equationautorefname~#1\null{%
	(#1)\null
}   
\xpretocmd{\eqref}{equation\,}{}{}
\usepackage{authblk}

\title{\textbf{\Large Active particle dynamics in uniform and patterned electroosmotic flows}}
\author[1]{Aritra Roy}
\author[2]{Antarip Poddar\thanks{Email: antarip@iitism.ac.in}$^{,}$}
\author[3]{Rahil Valani\thanks{Email: rahil.valani@physics.ox.ac.uk}$^{,}$}
\affil[1]{Department of Mechanical Engineering, Indian Institute of Technology Bombay, Powai, Mumbai 400076, India}
\affil[2]{Department of Mechanical Engineering, {Indian Institute of Technology (Indian School of Mines) Dhanbad}, Jharkhand 826004, India}
\affil[3]{Rudolf Peierls Centre for Theoretical Physics,
Parks Road,
University of Oxford,
Oxford OX1 3PU,
United Kingdom}

\date{}                     
\begin{document}
	\maketitle
	\linespread{1.1}
    \begin{abstract}
		\noindent
We investigate the dynamics of self-propelled spheroidal particles in electroosmotic channel flows generated by uniform, patterned, and combined zeta-potential distributions. The coupling between self-propulsion, particle shape, and electroosmotic advection gives rise to distinct regimes of confinement and transport. For a uniform zeta potential, the dynamics reduce to a two-dimensional system in cross-stream position and orientation. We show that this system possesses a non-canonical Hamiltonian structure, allowing its phase-space organization to be characterized analytically. For spherical particles, the dynamics admit a nonlinear-pendulum description, yielding the characteristic oscillation frequency and extent of the confined region. The oscillation frequency depends non-monotonically on electrostatic screening and reaches a maximum at an intermediate Debye parameter. Spatially patterned zeta potentials reorganize the dynamics by generating counter-rotating vortices and coupling streamwise position, cross-stream position, and particle orientation. Numerical simulations and dynamical-systems analysis reveal regimes of single-cell confinement and cross-cell migration, controlled by particle shape and electrostatic screening. When uniform and patterned components coexist, mean electroosmotic advection competes with vortex-induced recirculation, producing transport sensitive to particle shape and initial conditions. Together, these results connect Hamiltonian phase-space dynamics in uniform electroosmotic flow to vortex-mediated trapping and transport in patterned flows, highlighting the potential of surface electrostatic patterning to control particle motion without altering channel geometry.

\end{abstract}

\section{Introduction}
\label{sec:introduction}

\begin{figure}[t]
	\centering
    \includegraphics[width=1.03\textwidth]{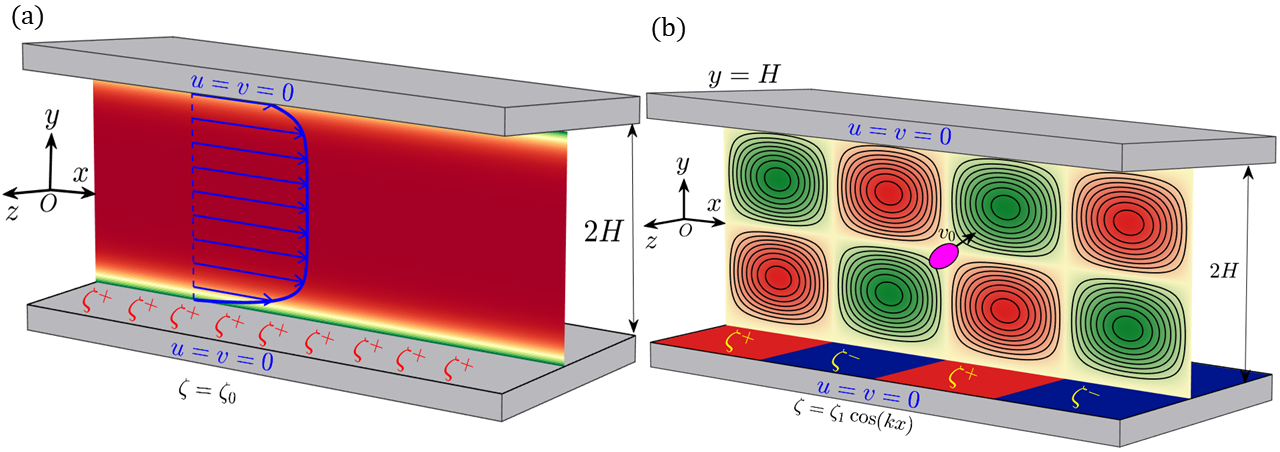}    
\caption{
Schematic of electroosmotic flow within a parallel-plate microfluidic channel bounded by solid walls at $y=\pm H$. 
(a) For a uniform zeta potential, $\phi(x,\pm H)=\zeta_0$, represented by a uniformly charged surface, the electroosmotic forcing generates a steady unidirectional flow along the channel. 
(b) For a periodically varying zeta potential, $\phi(x,\pm H)=\zeta_0\cos(kx)$, represented by alternating positive and negative surface charge regions, the spatially modulated electroosmotic forcing produces a steady array of counter-rotating cellular vortices, visualized by the red and green velocity contours and streamlines. A self-propelled spheroidal particle (purple), shown in panel (b), is suspended within the fluid and self-propels with an intrinsic speed $v_0$. In both configurations, the fluid velocity satisfies the no-slip and no-penetration boundary conditions, $u=v=0$, at the solid--liquid interfaces.}
	\label{fig:schematic}
\end{figure}

The transport and manipulation of suspended particles in microfluidic environments underpin a wide range of applications, including particle sorting, chemical processing, biomedical analysis, and lab-on-a-chip technologies \citep{stone2004engineering,whitesides2006origins,haeberle2007microfluidic}. At these length scales, hydrodynamic and externally imposed forcing mechanisms can be exploited to focus, sort, and manipulate suspended particles \citep{xuan2010particle,karimi2013hydrodynamic}. Electrokinetic methods are particularly attractive because they enable fluid transport without mechanically moving components and can be readily controlled through applied electric fields and surface properties \citep{rice1965electrokinetic,kirby2004zeta,chang2010electrokinetically}. Among these mechanisms, electroosmotic flow arises from the interaction between an applied electric field and the electrical double layer adjacent to a charged solid--liquid interface \citep{masliyah2006electrokinetic,hunter2013zeta}. For uniformly charged surfaces, this interaction produces the familiar unidirectional electroosmotic flow in a straight channel (figure~\ref{fig:schematic}(a)).

A richer class of flows emerges when the electrostatic properties of the bounding surfaces vary spatially. In particular, variations in the zeta potential can generate transverse motion, recirculation,
and cellular flow structures (figure~\ref{fig:schematic}b), even in geometrically simple channels \citep{Ajdari1995,herr2000electroosmotic,
ghosal2002lubrication,paratore2019}. More broadly, the sensitivity of electrokinetic transport to surface properties, including slip-dependent zeta potentials,
has also been demonstrated in the context of electrokinetic energy harvesting \citep{roy2026electrokinetic}. Such control over surface electrostatics provides a means of
modifying microscale transport without altering the channel geometry \citep{erickson2002influence,qian2002chaotic}. More recently, spatial and temporal modulation of the zeta potential has enabled dynamically reconfigurable microscale flow patterns \citep{mruetusatorn2009low,van2006directional,
paratore2019dynamic}. These developments raise the broader question of how structured electroosmotic flows interact with particles possessing their own intrinsic dynamics.

Active particles, which consume energy to generate persistent motion, provide a natural setting in which to address this question. Examples include biological microswimmers, such as bacteria and motile algae, as well as synthetic microswimmers, including catalytic Janus particles, light-driven colloids and microrobots \citep{elgeti2015physics,bechinger2016active}. Unlike passive tracers, these particles self-propel relative to the surrounding fluid, and their trajectories are therefore governed by the competition between autonomous propulsion and advection by the background flow. For non-spherical particles, the dynamics is further enriched by flow-induced reorientation, which depends on the local vorticity and rate of strain \citep{jeffery1922motion}. Consequently, even steady flows can generate non-trivial dynamics in the combined space of particle position and orientation. Previous studies of microswimmers in shear and channel flows, vortices, and periodic cellular flows have revealed oscillatory and rotational motion, trapping and escape, and transitions between localized and extended transport \citep{zottl2012nonlinear,torney2007transport,
tanasijevic2022microswimmers,Valani2024,Truneh2026-bu,Solomon2025}. Dynamical-systems approaches have proved particularly useful for understanding these behaviours, as fixed points, separatrices, conserved quantities, and bifurcations provide a framework for identifying the structures that organize particle motion \citep{zottl2012nonlinear,torney2007transport,Valani2024,Harding2025}.

The coupling between active propulsion and electroosmotic flow introduces additional, experimentally accessible means of controlling particle motion. Electrostatic screening and surface patterning govern the structure of the background flow, while swimming speed and particle geometry determine how particles respond to it.  Although active particles have been studied extensively in externally imposed shear, channel, and vortical flows, their behaviour in
electroosmotic flows remains comparatively unexplored. Of particular interest is the ability to continuously vary the relative contributions of uniform and patterned zeta potentials, thereby connecting predominantly unidirectional advection to vortex-dominated cellular motion. Understanding how particle trajectories and transport evolve across these regimes provides the central motivation for the present study.

In this work, we investigate the dynamics of self-propelled spheroidal
particles in electroosmotic channel flows generated by uniform, spatially
patterned, and combined zeta-potential distributions. We first derive
the background electroosmotic flow and formulate the coupled particle
dynamics in \S\ref{sec:formulation}. For a uniform zeta potential,
\S\ref{sec:uniform} establishes a non-canonical Hamiltonian description
of particle confinement and oscillatory motion, with a nonlinear-pendulum
limit for spherical particles. In contrast, spatially patterned zeta
potentials generate cellular vortices and fully coupled positional and
orientational dynamics. In \S\ref{sec:patterned}, numerical trajectories
and bifurcation diagrams reveal how particle shape and electrostatic
screening control single-cell confinement and cross-cell transport.
We then examine the intermediate regime in \S\ref{sec:Combined}, where
uniform advection and vortex-induced recirculation compete to determine
particle transport and its sensitivity to initial conditions. Finally,
\S\ref{sec:summary} summarizes the principal findings and their
implications for controlling active particle transport through
electroosmotic surface patterning.

\section{Mathematical formulation}
\label{sec:formulation}
A schematic of the system is shown in Fig.~\ref{fig:schematic}. We consider electroosmotic flow in a two-dimensional parallel-plate channel of height $2H$, bounded by solid walls located at $y=\pm H$. The channel is assumed to be invariant in the spanwise direction, such that the fluid motion is confined to the $(x,y)$-plane. {Upon contact with an electrolyte solution, the channel walls acquire a surface charge through the ionization of surface functional groups or the adsorption of ions. This surface charge modifies the local ionic distribution, attracting counterions and repelling co-ions. The resulting electrostatic redistribution of ions is opposed by thermal diffusion, which tends to restore a uniform concentration. The interplay between these effects establishes an electrical double layer (EDL) adjacent to each wall, comprising a compact Stern layer and a diffuse layer of mobile ions. The electrostatic potential at the shear plane, conventionally approximated as the outer edge of the Stern layer, is characterized by the zeta potential, $\zeta$~\citep{stern1924,grahame1947,
kirby2004zeta,masliyah2006electrokinetic}. 
In the bulk region, sufficiently far from the walls, the electrolyte is approximately electrically neutral. Upon application of an external electric field $E_x$ along the channel, the net charge density $\rho_e$ within the EDL experiences an electrical body force per unit volume, $\rho_e E_x$. The resulting motion of the mobile ions transfers momentum to the surrounding liquid through viscous interactions, thereby generating electroosmotic flow~\citep{rice1965electrokinetic,
masliyah2006electrokinetic}.}
Figure~\ref{fig:schematic}(a) illustrates the case of a uniform
zeta potential, which produces a unidirectional electroosmotic flow,
whereas figure~\ref{fig:schematic}(b) shows a spatially varying zeta
potential that generates a periodic array of counter-rotating vortices. To encompass both limits, as well as intermediate configurations in which
uniform and spatially varying components coexist, we prescribe the same
zeta-potential distribution along the upper and lower walls
\begin{equation}
    \zeta(x)=\zeta_0+\zeta_1\cos(kx),
     \label{eq:zeta_distribution}
\end{equation}
where $\zeta_0$ denotes the uniform component, $\zeta_1$ is the amplitude
of the spatial modulation, and $k$ is its wavenumber. The patterned
component has streamwise wavelength $2\pi/k$, and periodic boundary
conditions are imposed on the fluid variables over one spatial period.
A self-propelled spheroidal particle is suspended in the channel and
swims with intrinsic speed $v_0$ relative to the surrounding fluid. Its
motion results from the combined effects of self-propulsion, advection by
the electroosmotic flow, and flow-induced reorientation. The distribution of the electric potential $\phi$ within the electric double layer (EDL) is governed by the Poisson-Boltzmann (PB) equation, expressed in dimensional form as \citep{kirby2004zeta,masliyah2006electrokinetic}
\begin{equation}
\boldsymbol{\nabla}^2 \phi = -\frac{e}{\varepsilon} \sum_{i=1}^{m} z_i n_{\infty} \exp\left( -\frac{z_i e \phi}{k_B T} \right)
\label{eq:pb_general}
\end{equation}
where $e$ is the elementary charge, $k_B$ is the Boltzmann constant, $T$ is the absolute temperature, $n_{i,\infty}$ represents the bulk numerical concentration of the $i$-th ionic species with valence $z_i$, and $\varepsilon = \varepsilon_0 \varepsilon_r$ denotes the absolute permittivity of the electrolyte solution. Assuming a symmetric electrolyte ($z:z$) with a net charge density invariant with temperature and neglecting steric ion-size effects, (\ref{eq:pb_general}) simplifies by defining the individual ionic number densities via the Boltzmann equilibrium distribution, $n_\pm = n_\infty \exp\left(\mp z e \phi / k_B T\right)$. 
For a homogeneous, incompressible fluid, under the assumption of weak temperature dependence of the dielectric properties, the hydrodynamic motion is coupled to the electrostatic field via the electric body force per unit volume. This forcing term is represented by the divergence of the deviatoric stress tensor $\boldsymbol{\sigma}$ as \citep{Batchelor2000,jackson2012classical}
\begin{equation}
\boldsymbol{\nabla} \cdot \boldsymbol{\sigma} + \boldsymbol{f_e} = 0, \quad \left[\boldsymbol{\sigma} = - p\mathbb{I} + \mu(\boldsymbol{\nabla}\boldsymbol{u} + \boldsymbol{\nabla} \boldsymbol{u}^T) \right]
\label{eq:momentum_stress}
\end{equation}
where $\rho_e = z e (n_+ - n_-)$ defines the net volumetric charge density. Inserting this electric body force directly into the momentum equation provides the driving mechanism for electro-osmotic transport across the fluid domain. To systematically characterize the competing electrohydrodynamic forces and facilitate parametric analysis, these dimensional governing relations are subsequently cast into non-dimensional form.

We nondimensionalize the governing equations using the channel half-height
$H$ as the characteristic length scale and the Helmholtz--Smoluchowski
velocity based on a reference zeta potential, as the characteristic velocity scale, following the same methodloogy as \citep{arcos2018dispersion,roy2026electrokinetic}. The dimensionless variables, denoted temporarily by a superscript $*$,
and the principal dimensionless parameters are defined as
\begin{align*}
x^* &= \frac{x}{H}, \quad
y^* = \frac{y}{H}, \quad
\mathbf{u}^* = \frac{\mathbf{u}}{U_{eo}}, \quad
\phi^* = \frac{\phi}{\zeta_{ref}}, \quad 
\zeta_{\mathrm{ref}}=\zeta_0+\zeta_1
\\
k^* &= kH, \quad
K = \kappa H = \frac{H}{\lambda_D}, \quad
\eta = \frac{\zeta_1}{\zeta_{\mathrm{ref}}}, \quad
\lambda = \sqrt{K^2+k^{*2}}, \quad 
U_{eo}=-\frac{\varepsilon E_x\zeta_{\mathrm{ref}}}{\mu}
\end{align*}
Here, $k^*$ is the dimensionless wavenumber of the surface pattern, $\kappa^2 = 2n_{\infty} e^2 z^2 / (\varepsilon k_B T)$ is the Debye parameter, $K=H/\lambda_D$ is the dimensionless is electrokinetic separation distance, where
$\lambda_D=\kappa^{-1}$ denotes the electrical double layer thickness, and $\eta$
measures the relative contribution of the spatially varying component of
the zeta potential. Here, $E_x$ is
the externally applied electric field. Thus, $\eta=0$ corresponds to a uniform zeta potential, $\eta=1$ to a
purely sinusoidal distribution, and $0<\eta<1$ to a superposition of the
two. The quantity $\lambda$ is introduced for notational convenience in
the analytical solution below.

We next determine the electric-potential distribution within the channel
resulting from the prescribed wall zeta potential. Assuming a dilute,
symmetric electrolyte and sufficiently small electrostatic potentials, the
Poisson--Boltzmann equation may be linearized using the Debye--H\"uckel
approximation~\citep{masliyah2006electrokinetic,hunter2013zeta}. This assumption is justified since the electrostatic potential energy is significantly smaller than thermal kinetic energy ($\vert{}z e \psi\vert{} \ll k_B T$), restricting its validity to low potential regimes below approximately $25\text{ mV}$. Physically, this holds because random thermal agitation dominates weak electrostatic forces near weakly charged surfaces \citep{kirby2004zeta,chakraborty2008streaming,roy2026electrokinetic}. The dimensionless boundary condition subject to the wall condition $\phi(x,\pm1)=(1-\eta)+\eta\cos(kx)$, and the centreline symmetry condition
$\partial\phi / \partial y|_{y=0}=0$. Here and throughout the remainder of the paper, the superscript $*$ is
omitted for notational simplicity. Owing to the linearity of the
Debye--H\"uckel equation, the potential can be decomposed into a uniform
component and a spatially periodic component proportional to $\cos(kx)$.
Solving the resulting dimensionless form of the ordinary differential equations (\ref{eq:pb_general}) in $y$ and applying
the wall and centreline symmetry conditions, respectively, gives
\begin{equation}
\phi(x,y)
=
(1-\eta)\frac{\cosh(Ky)}{\cosh K}
+
\eta\frac{\cosh(\lambda y)}{\cosh\lambda}\cos(kx),
\label{eq:epot_sol}
\end{equation}
where $\lambda=\sqrt{K^2+k^2}$. In the low-Reynolds-number limit, the fluid motion is governed by the
Stokes equations with the electrostatic body force associated with the
diffuse charge in the electrical double layer. Introducing the
streamfunction $\Psi$ \citep{Chang2008,Morozov2018,wang2023electroosmotic}, such that $u=\partial\Psi/\partial y$, and $v=-\partial\Psi/\partial x$, satisfies incompressibility condition,
\begin{equation}
\boldsymbol{\nabla}\cdot\boldsymbol{u}=0
\label{eq:cont_dim}
\end{equation}
identically. The momentum equation has an electric forcing on the flow field denoted using $\boldsymbol{f_e} = \rho_e E_x \hat{\mathbf{i}}$ \citep{Ajdari1995,Morozov2018}. Eliminating the pressure by taking the curl of the momentum
equation (\ref{eq:momentum_stress}) gives
\begin{equation}
\nabla^{4}\Psi = -\frac{\partial(\nabla^{2}\phi)}{\partial y} 
\label{eq:mom_dim}
\end{equation}
Substituting (\ref{eq:epot_sol}) into (\ref{eq:mom_dim}) and imposing
no-slip and no-penetration at the channel walls, together with centreline
symmetry and streamwise periodicity, yields
\begin{equation}
\Psi(x,y)
=
(1-\eta)
\left[
y-\frac{\sinh(Ky)}{K\cosh K}
\right]
+
\eta
\left[
B\sinh(ky)
+Cy\cosh(ky)
-\frac{\lambda\sinh(\lambda y)}
{K^2\cosh\lambda}
\right]\cos(kx).
\label{eq:hydrosol_1}
\end{equation}
The constants $B$ and $C$, together with the detailed derivation of
(\ref{eq:hydrosol_1}), are given in Appendix~\ref{appenA}.
\subsection{Active particle dynamics}
The translational and orientational dynamics of a self-propelled spheroidal particle advected by the electroosmotic flow are described by \citep{jeffery1922motion,torney2007transport,Solomon2025}
\begin{subequations} 
\begin{align}
\frac{\mathrm{d}\boldsymbol{r}_p}{\mathrm{d}t}
&=
\boldsymbol{u}(\boldsymbol{r}_p,t)+v_0\boldsymbol{p}
\label{eq:eom_pos}\\
\frac{\mathrm{d}\boldsymbol{p}}{\mathrm{d}t}
&=
\frac{1}{2}\boldsymbol{\omega}\times\boldsymbol{p}
+
\alpha\,\boldsymbol{p}\cdot\boldsymbol{E}\cdot
(\boldsymbol{I}-\boldsymbol{p}\boldsymbol{p})
\label{eq:eom_ori}
\end{align}
\end{subequations}
where $\boldsymbol{r}_p=(x_p,y_p)$ denotes the particle position, $v_0$ is its intrinsic swimming speed, and
$\boldsymbol{p}=(\cos\theta_p,\sin\theta_p)$ is the unit vector along its swimming direction. The background fluid velocity is
$\boldsymbol{u}=(u,v)$, with vorticity
$\boldsymbol{\omega}=\boldsymbol{\nabla}\times\boldsymbol{u}$
and rate-of-strain tensor $\boldsymbol{E}
=
\frac{1}{2}
\left(
\boldsymbol{\nabla}\boldsymbol{u}
+
\boldsymbol{\nabla}\boldsymbol{u}^{T}
\right)$ while $\boldsymbol{I}$ denotes the identity tensor. The particle shape is characterized by $\alpha=(\gamma^2-1)/(\gamma^2+1)$ where $\gamma$ is the particle aspect ratio. The model assumes that the active particle has a negligible electrophoretic mobility. This condition may be realized when the particle is nearly uncharged \citep{OBrien2014electrophoretic} or when its zeta potential is close to zero, at its isoelectric point ($pI$) \citep{Knecht2008}, applying specific uncharged surface functionalization, or using high-salt buffer solutions where the particle zeta potential $\zeta_p \ll \zeta_{wall}$). Under this assumption, electrophoretic migration relative to the suspending fluid is neglected ($\mu_{EP} \approx 0$), and  the applied electric field acts exclusively on the electric double layer established along the surrounding substrate and bulk fluid medium, generating an electroosmotic flow that hydrodynamically advects the particle. Using the streamfunction representation of the background flow and writing the particle orientation in terms of $\theta_p$, equations~(\ref{eq:eom_pos})--(\ref{eq:eom_ori}) reduce to
\begin{subequations}
\begin{align}
\frac{\mathrm{d}x_p}{\mathrm{d}\tau}
&=
\frac{\partial\Psi}{\partial y_p}
+
v_s\cos\theta_p
\label{eq:dxdt_spheroid}\\
\frac{\mathrm{d}y_p}{\mathrm{d}\tau}
&=
-\frac{\partial\Psi}{\partial x_p}
+
v_s\sin\theta_p
\label{eq:dydt_spheroid}\\
\frac{\mathrm{d}\theta_p}{\mathrm{d}\tau}
&=
-\frac{1}{2}\nabla^2\Psi
+
\alpha
\left[
\frac{1}{2}
\left(
\frac{\partial^2\Psi}{\partial y_p^2}
-
\frac{\partial^2\Psi}{\partial x_p^2}
\right)
\cos(2\theta_p)
-
\frac{\partial^2\Psi}{\partial x_p\partial y_p}
\sin(2\theta_p)
\right]
\label{eq:dthetadt_spheroid}
\end{align}
\end{subequations}
Here, $v_s=v_0/U_{eo}$ and $\tau=U_{eo}t/H$ denote the
dimensionless swimming speed and time, respectively.
To account for particle--wall interactions, we impose a
simplified steric no-penetration condition at
$y_p=\pm(1-\epsilon)$, where $\epsilon (= 10^{-5}) \ll1$ is a small
dimensionless offset from the channel walls. When a particle reaches either boundary with an outward wall-normal velocity, its cross-stream position is pinned at the boundary, while tangential motion remains unrestricted. The particle orientation continues to evolve under the local hydrodynamic vorticity and strain, without an imposed reflection of its swimming direction. Once the wall-normal component of its
velocity points towards the channel interior, the particle is released and resumes its unconstrained motion. This idealized wall-pinning condition prevents particle penetration without explicitly resolving near-wall hydrodynamic or steric interactions \citep{elgeti2013wall}.

\section{Hamiltonian active-particle dynamics in constant zeta potential}
\label{sec:uniform}
We first consider the case of a uniform zeta potential, $\eta=0$. Hence, the streamfunction in equation~(\ref{eq:hydrosol_1}) is independent of the
streamwise coordinate $x$, and the electroosmotic flow is unidirectional,
with the fluid velocity varying only across the channel width (figure~\ref{fig:schematic}(a)). Consequently,
the streamwise position $x_p$ does not influence the evolution of the
cross-stream position $y_p$ or the particle orientation $\theta_p$.
The dynamics therefore reduce to the following  two-dimensional  autonomous dynamical system
\begin{subequations}
\begin{align}
\frac{\mathrm{d}y_p}{\mathrm{d}\tau}
&=
v_s\sin\theta_p,
\label{eq:dy_1}\\
\frac{\mathrm{d}\theta_p}{\mathrm{d}\tau}
&=
-\frac{K}{2}
\frac{\sinh(Ky_p)}{\cosh K}
\left[
\alpha\cos(2\theta_p)-1
\right],
\label{eq:dtheta_1}
\end{align}
\end{subequations}
where $\tau$ is the dimensionless time. Thus, the cross-stream position and particle orientation form a closed dynamical system, while the streamwise trajectory can be determined independently once $(y_p,\theta_p)$ is known. The equilibrium states $(y_p^*,\theta_p^*)$ are obtained by setting both right-hand sides  equations~(\ref{eq:dy_1})--(\ref{eq:dtheta_1}) to zero. For an active particle with $v_s>0$, Eq.~(\ref{eq:dy_1}) requires $\sin\theta_p^*=0$. Hence, the possible steady orientations are $\theta_p^*=0$ and $\theta_p^*=\pi$, modulo $2\pi$. Moreover, within the physical range $-1<\alpha<1$, the factor $1-\alpha\cos(2\theta_p)$ remains strictly positive. Equation (\ref{eq:dtheta_1}) therefore requires $\sinh(Ky_p^*)=0$, which gives $y_p^*=0$. 
Hence, the reduced system possesses two distinct equilibrium states,
$(y_p^*,\theta_p^*)=(0,0)$ and $(0,\pi)$, both located at the channel
centreline. The former corresponds to a particle oriented downstream,
along the direction of the background electroosmotic flow, whereas
the latter represents a particle oriented upstream, against the flow.
At these states, the cross-stream swimming velocity vanishes and the
particle experiences no net hydrodynamic reorientation, so that its
cross-stream position and orientation remain unchanged.

To determine the local dynamics near these equilibria, we linearize
(\ref{eq:dy_1})--(\ref{eq:dtheta_1}). The corresponding Jacobian is
\begin{equation}
\mathcal{J}(y_p,\theta_p)
=
\begin{bmatrix}
0
&
v_s\cos\theta_p
\\[4pt]
-\dfrac{K^2}{2}
\dfrac{\cosh(Ky_p)}{\cosh K}
\left[
\alpha\cos(2\theta_p)-1
\right]
&
K\alpha
\dfrac{\sinh(Ky_p)}{\cosh K}
\sin(2\theta_p)
\end{bmatrix}.
\label{eq:matrix}
\end{equation}
Evaluating the characteristic equation at the equilibria gives
\begin{equation}
\lambda^2
=
\frac{v_sK^2(1-\alpha)}
     {2\cosh K}
\cos\theta_p^*.
\label{eq:eigen_general}
\end{equation}
The two equilibria therefore have eigenvalues
\begin{equation}
\left.\lambda_{1,2}\right|_{(0,0)}
=
\pm
\sqrt{
\frac{v_sK^2(1-\alpha)}
     {2\cosh K}
},
\quad
\left.\lambda_{1,2}\right|_{(0,\pi)}
=
\pm i
\sqrt{
\frac{v_sK^2(1-\alpha)}
     {2\cosh K}
}.
\end{equation}
For $v_s>0$ and $-1<\alpha<1$, the equilibrium at $(0,0)$ is therefore
hyperbolic and corresponds to a saddle, whereas $(0,\pi)$ is elliptic
and corresponds to a centre. Small perturbations about the former grow
along its unstable direction, while trajectories sufficiently close to
the latter undergo bounded oscillations in cross-stream position and
orientation. These two equilibria consequently organize the local
phase-space dynamics of the particle.
The coexistence of an elliptic centre, a hyperbolic saddle, and closed
phase-space trajectories suggests that the reduced system possesses an
underlying conservative structure. For a canonical Hamiltonian system
with coordinates $(y_p,\theta_p)$, the equations of motion would take the
form
\begin{equation}
\dot{y}_p
=
\frac{\partial H}{\partial\theta_p},
\qquad
\dot{\theta}_p
=
-\frac{\partial H}{\partial y_p}
\label{eq:canonical_Hamilton}
\end{equation}
and the corresponding vector field would be divergence-free in the
$(y_p,\theta_p)$ phase plane. Equations~(\ref{eq:dy_1})--(\ref{eq:dtheta_1})
do not satisfy this condition for $\alpha\neq0$, and hence
$(y_p,\theta_p)$ are not canonical coordinates. The conservative structure can nevertheless be recovered by introducing
an integrating factor, or equivalently an invariant phase-space density
$\mu(\theta_p)$, such that the weighted vector field becomes
divergence-free. As shown in Appendix~\ref{appenB}, the required density is
\begin{equation}
\mu(\theta_p)
=
\frac{1}{1-\alpha\cos(2\theta_p)}.
\label{eq:integrating_factor}
\end{equation}
The equations of motion can then be written in the non-canonical
Hamiltonian form
\begin{equation}
\dot{y}_p
=
\frac{1}{\mu(\theta_p)}
\frac{\partial H}{\partial\theta_p},
\qquad
\dot{\theta}_p
=
-\frac{1}{\mu(\theta_p)}
\frac{\partial H}{\partial y_p}.
\label{eq:noncanonical_Hamilton}
\end{equation}
Thus, the system retains a Hamiltonian description, but with a
state-dependent prefactor $1/\mu(\theta_p)$ multiplying the canonical
Hamiltonian vector field. Equivalently, the dynamics preserves the
weighted phase-space measure
$\mu(\theta_p)\,\mathrm{d}y_p\,\mathrm{d}\theta_p$, rather than the
ordinary area element $\mathrm{d}y_p\,\mathrm{d}\theta_p$. The departure
from canonical form therefore arises from the shape-dependent
reorientation of a non-spherical particle; in the spherical limit
$\alpha=0$, $\mu=1$ and the standard canonical Hamilton equations are
recovered.

Integrating (\ref{eq:noncanonical_Hamilton}) gives the conserved
Hamiltonian (see Appendix~\ref{appenB})
\begin{equation}
H(y_p,\theta_p)
=
-\frac{\cosh(Ky_p)}{2\cosh K}
+
\begin{cases}
\displaystyle
-\frac{v_s}{\sqrt{-2\alpha(1+\alpha)}}
\arctan\left(
\sqrt{\frac{-2\alpha}{1+\alpha}}\,
\cos\theta_p
\right),
& -1<\alpha<0,
\\[12pt]
\displaystyle
-v_s\cos\theta_p,
& \alpha=0,
\\[12pt]
\displaystyle
-\frac{v_s}{\sqrt{2\alpha(1+\alpha)}}
\operatorname{artanh}\left(
\sqrt{\frac{2\alpha}{1+\alpha}}\,
\cos\theta_p
\right),
& 0<\alpha<1.
\end{cases}
\label{eq:hamiltonian_general_m}
\end{equation}
Along a particle trajectory,
\begin{equation}
\frac{\mathrm{d}H}{\mathrm{d}\tau}
=
\frac{\partial H}{\partial y_p}\dot{y}_p
+
\frac{\partial H}{\partial\theta_p}\dot{\theta}_p
=0,
\label{eq:H_conservation}
\end{equation}
so that particle trajectories lie on level sets of $H$ in the
$(y_p,\theta_p)$ phase plane. The Hamiltonian therefore provides a direct
description of the global phase-space organization where closed level sets
surrounding the elliptic centre correspond to bounded oscillations in
cross-stream position and orientation, while the separatrix passing
through the hyperbolic saddle divides these oscillatory trajectories from
the surrounding rotational dynamics.

\begin{figure}[t]
    \centering
    \includegraphics[width=1.08\textwidth]{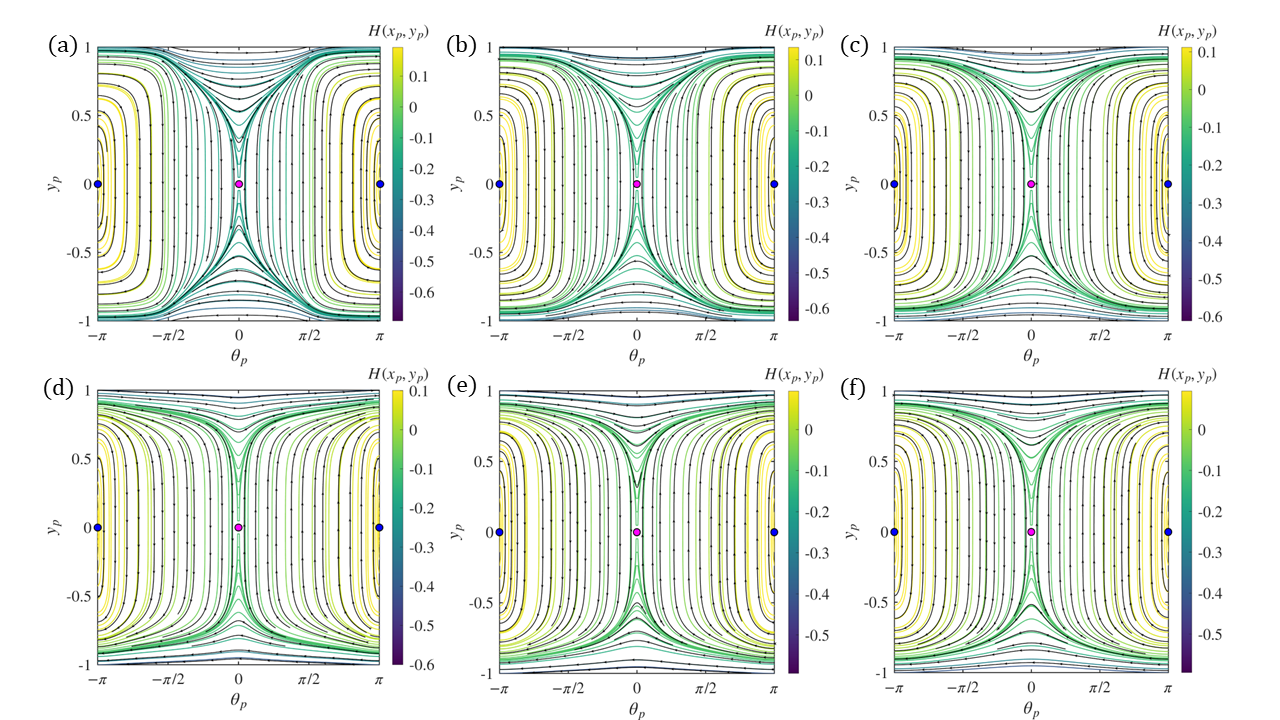}
    \caption{Phase-space trajectories (black directed curves) superimposed
on contours of the Hamiltonian $H(y_p,\theta_p)$ for $v_s=0.1$ and
$K=10$. Panels ($a$--$c$) and ($d$--$f$) show representative particle
shapes with $\alpha=-0.75,-0.50,-0.25$ and
$\alpha=0.75,0.50,0.25$, respectively. Markers denote the hyperbolic
saddle at $(y_p,\theta_p)=(0,0)$ (\magenmarker) and the elliptic centres
at $(0,\pm\pi)$ (\bluemarker). The coincidence between particle
trajectories and constant-$H$ contours illustrates conservation of the
Hamiltonian, while variations in $\alpha$ modify the geometry and
cross-stream extent of the separatrix.}
    \label{fig:pp_ham}
\end{figure}

The Hamiltonian structure is illustrated in
figure~\ref{fig:pp_ham}, where trajectories obtained from the dynamical
system are superimposed on contours of the Hamiltonian in
\eqref{eq:hamiltonian_general_m}. The trajectories closely follow the contours for all values of $\alpha$ considered, providing a
direct numerical illustration of Hamiltonian conservation. The
saddle--centre topology is preserved as the particle shape is varied,
with the hyperbolic equilibrium at $(0,0)$ and the elliptic equilibria at
$(0,\pi)$. Particle anisotropy nevertheless modifies the geometry of
the level sets and, in particular, the shape and cross-stream extent of
the separatrix. Thus, while $\alpha$ does not alter the qualitative
saddle--centre organization of the uniform-zeta dynamics, it controls the
phase-space region occupied by bounded oscillatory trajectories.

\begin{figure}[t]
    \centering
    \includegraphics[width=1.04\textwidth]{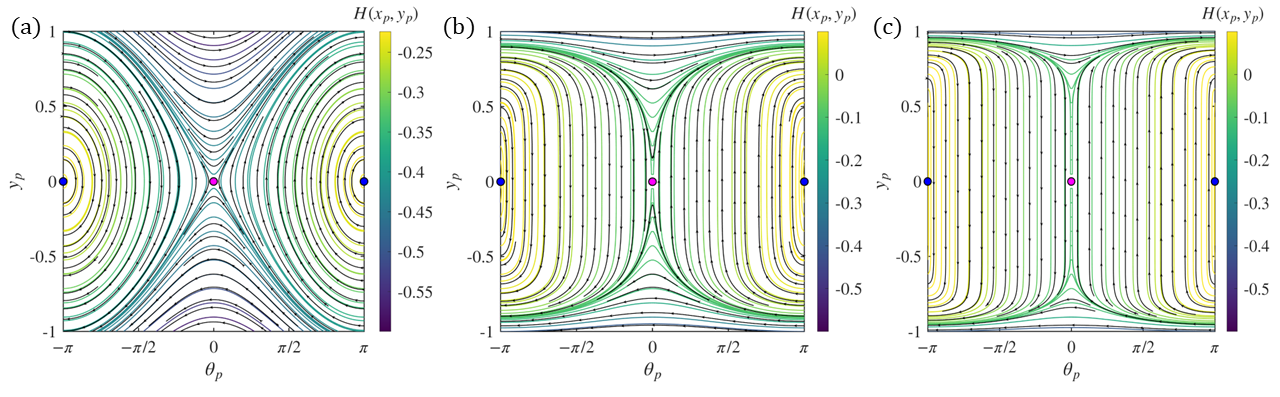}
   \caption{Phase-space trajectories superimposed on Hamiltonian contours
$H(y_p,\theta_p)$ for ($a$) $K=1$, ($b$) $K=10$, and ($c$) $K=20$,
with $v_s=0.1$ and $\alpha=0$. The hyperbolic saddle at
$(y_p,\theta_p)=(0,0)$ (\magenmarker) and elliptic centres at
$(0,\pm\pi)$ (\bluemarker) organize the phase space. At $K=1$, the
system exhibits the characteristic cat-eye topology of a nonlinear
pendulum. Increasing $K$ preserves the saddle--centre structure while
progressively modifying the geometry and cross-stream extent of the
oscillatory region, illustrating the influence of electrostatic
screening on particle confinement.}
  \label{fig:phase_portrait_Kvary}
\end{figure}
\begin{figure}[t]
    \centering
    \includegraphics[width=1.02\textwidth]{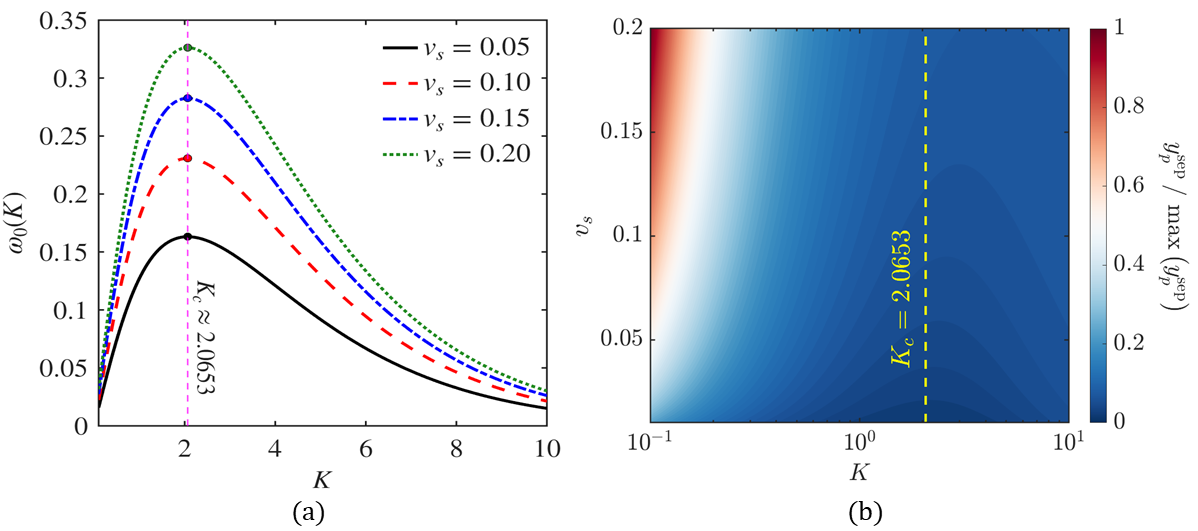}
   \caption{Influence of the Debye parameter $K$ and swimming speed $v_s$
on the dynamics and phase-space confinement of a spherical active
particle. ($a$) Small-amplitude oscillation frequency $\omega_0$ as a
function of $K$ for $v_s=0.05$, $0.10$, $0.15$, and $0.20$. The
frequency reaches its maximum at $K_c\simeq2.0653$ (vertical dashed
line), determined by $K_c\tanh K_c=2$, independently of $v_s$.
($b$) Normalized maximum separatrix half-width
$y_{p,\max}^{\mathrm{sep}}/\max(y_{p,\max}^{\mathrm{sep}})$ over the
$(K,v_s)$ parameter space, with $K$ shown on a logarithmic scale. The
two panels characterize, respectively, the time scale and cross-stream
extent of oscillatory particle motion.}
  \label{fig:freq_sepatrix}
\end{figure}

\subsection{The nonlinear pendulum limit for a spherical active particle}
\label{sec:pendulum}

We now consider the limit of a spherical active particle, $\alpha=0$. In this case,
the shape-dependent contribution to the orientational dynamics vanishes
and the integrating factor introduced in the previous section reduces to
$\mu=1$. The non-canonical Hamiltonian system therefore becomes canonical,
with Hamiltonian
\begin{equation}
H(y_p,\theta_p)
=
-\frac{\cosh(Ky_p)}{2\cosh K}
-v_s\cos\theta_p.
\label{eq:hamilreduce}
\end{equation}

The phase-space dynamics resembles that of a nonlinear pendulum, with
$\theta_p$ and $y_p$ serving as the two phase-space variables.
In this analogy,
$(y_p,\theta_p)=(0,\pi)$ corresponds to the stable pendulum equilibrium
and is surrounded by closed oscillatory trajectories, representing
bounded oscillations of the particle about the channel centreline.
Conversely, $(0,0)$ corresponds to the unstable equilibrium and forms a
hyperbolic saddle in the phase plane. The separatrix passing through this
saddle divides the closed oscillatory trajectories from rotational
trajectories, thereby providing a natural boundary between distinct
classes of particle motion.


The influence of electrostatic screening on this phase-space structure is
shown in figure~\ref{fig:phase_portrait_Kvary}. For all values of $K$
considered, the saddle--centre topology and the associated separatrix are
preserved, but the geometry of the oscillatory region changes markedly.
For $K=1$, the phase portrait displays the familiar cat-eye structure of
a nonlinear pendulum, with closed oscillatory orbits surrounding the
elliptic centres. As $K$ increases, the separatrix becomes increasingly
elongated in the cross-stream direction and the local curvature of the
Hamiltonian near the elliptic centre changes. This variation with $K$
suggests that electrostatic screening controls both the characteristic
time scale and the spatial extent of the confined particle motion. These
two effects are quantified below through the small-amplitude oscillation
frequency and the separatrix width.

\subsubsection{Small-amplitude dynamics about the elliptic centre}

To characterize the local dynamics near the stable elliptic centre, we
consider small perturbations about
$(y_p^*,\theta_p^*)=(0,\pi)$ and write $y_p=\widetilde{y}_p$, and $\theta_p=\pi+\widetilde{\theta}_p$ with $|\widetilde{\theta}_p|\ll1$,
and $K|\widetilde{y}_p|\ll1$. Under these conditions, we can invoke the small angle approximation that results in $\sin(\pi+\widetilde{\theta}_p)
\simeq-\widetilde{\theta}_p$, and $\sinh(K\widetilde{y}_p)
\simeq K\widetilde{y}_p$. Upon substituting these approximations into
equations~(\ref{eq:dy_1})--(\ref{eq:dtheta_1}) yields the
linearized system
\begin{subequations}
\label{eq:reddynamic}
\begin{align}
\dot{\widetilde{\theta}}_p
&\simeq
\frac{K^2}{2\cosh K}\,
\widetilde{y}_p,
\label{eq:reddynamic_a}\\
\dot{\widetilde{y}}_p
&\simeq
-v_s\widetilde{\theta}_p.
\label{eq:reddynamic_b}
\end{align}
\end{subequations}
Differentiating (\ref{eq:reddynamic_a}) with respect to the dimensionless
time $\tau$ and using (\ref{eq:reddynamic_b}) gives
\begin{equation}
\ddot{\widetilde{\theta}}_p
+
\omega_0^2\widetilde{\theta}_p
=0,
\qquad
\omega_0^2
=
\frac{v_sK^2}{2\cosh K}.
\label{eq:natural_frequency}
\end{equation}
Thus, sufficiently close to the elliptic centre, the nonlinear particle
dynamics reduces to that of a simple harmonic oscillator, where the particle orientation executes small-amplitude oscillatory dynamics known as swinging motion~\citep{zottl2012nonlinear}. The corresponding
oscillation frequency,
\begin{equation}
\omega_0
=
\sqrt{\frac{v_sK^2}{2\cosh K}},
\end{equation}
sets the characteristic time scale of small-amplitude oscillations in
cross-stream position and orientation. It increases as $\sqrt{v_s}$ with
the swimming speed, whereas its dependence on the Debye parameter $K$ is
non-monotonic. The value of $K$ that maximizes the oscillation frequency is obtained by
setting $\mathrm{d}\omega_0^2/\mathrm{d}K=0$, which gives
a transcendental equation $K_c\tanh K_c=2$. The positive solution is $K_c\simeq2.0653$.
The small-amplitude oscillation frequency therefore reaches a maximum at an
intermediate Debye parameter, rather than varying monotonically with the
EDL thickness. 

\subsubsection{Separatrix and extent of the oscillatory region}

The small-amplitude analysis above describes the dynamics only in the
neighbourhood of the elliptic centre. To characterize the global
phase-space structure, we now return to the full nonlinear Hamiltonian
(\ref{eq:hamilreduce}). As shown in
figure~\ref{fig:phase_portrait_Kvary}, the closed oscillatory trajectories
surrounding the elliptic centre are bounded by a separatrix that passes
through the hyperbolic saddle at $(y_p,\theta_p)=(0,0)$. This separatrix
therefore defines the boundary between bounded oscillatory motion (swinging motion) and
rotational trajectories in the $(y_p,\theta_p)$ phase plane, known as tumbling motion~\citep{zottl2012nonlinear}.

Since the Hamiltonian is conserved along particle trajectories, the
separatrix corresponds to the Hamiltonian level passing through the
hyperbolic saddle. Its value is
\begin{equation}
H_s
=
H(0,0)
=
-\frac{1}{2\cosh K}-v_s.
\label{eq:Hs}
\end{equation}
Hence, points $(y_p,\theta_p)$ lying on the separatrix satisfy
\begin{equation}
-\frac{\cosh(Ky_p)}{2\cosh K}
-v_s\cos\theta_p
=
-\frac{1}{2\cosh K}
-v_s.
\label{eq:separatrix_balance}
\end{equation}
Rearranging gives
\begin{equation}
\cosh(Ky_p)
=
1+2v_s\cosh K\left(1-\cos\theta_p\right).
\end{equation}
Solving for $y_p$ yields the upper and lower branches of the separatrix,
\begin{equation}
y_p^{\mathrm{sep}}(\theta_p)
=
\pm\frac{1}{K}
\operatorname{arcosh}
\left[
1+
4v_s\cosh K
\sin^2\left(\frac{\theta_p}{2}\right)
\right].
\label{eq:separatrix}
\end{equation}

Equation~(\ref{eq:separatrix}) gives the boundary of the oscillatory
region (swinging motion) explicitly in terms of the swimming speed and Debye parameter.
The largest cross-stream excursion of the separatrix occurs at
$\theta_p=\pi$, where $\sin^2(\theta_p/2)=1$. The corresponding
separatrix half-width is therefore
\begin{equation}
y_{p,\max}^{\mathrm{sep}}
=
\frac{1}{K}
\operatorname{arcosh}
\left(
1+4v_s\cosh K
\right).
\label{eq:separatrix_width}
\end{equation}
This quantity provides a convenient measure of the cross-stream extent of
the oscillatory region. Together with the small-amplitude frequency
derived above, it characterizes two complementary aspects of confinement:
 (1) $\omega_0$ determines the characteristic time scale of motion near the
elliptic centre, whereas (2) $y_{p,\max}^{\mathrm{sep}}$ measures the spatial
extent of the surrounding oscillatory region. 

The combined influence of electrostatic screening and swimming activity is
summarized in figure~\ref{fig:freq_sepatrix}. Figure~\ref{fig:freq_sepatrix}(a)
shows the small-amplitude oscillation frequency predicted by
(\ref{eq:natural_frequency}). For each swimming speed, $\omega_0$
increases with $K$ at small $K$, reaches a maximum at
$K=K_c\simeq2.0653$, and subsequently decreases as the EDL becomes progressively thinner. Increasing $v_s$ raises the
oscillation frequency through the scaling $\omega_0\propto\sqrt{v_s}$ but
does not shift the location of the maximum, since $K_c$ is determined
solely by $K_c\tanh K_c=2$.

Figure~\ref{fig:freq_sepatrix}(b) complements this local measure of the
dynamics by showing the maximum separatrix half-width obtained from
(\ref{eq:separatrix_width}) across the $(K,v_s)$ parameter space. The
separatrix width quantifies the cross-stream extent of the oscillatory
region (swinging motion) and therefore provides a global measure of phase-space confinement.
Taken together, the two panels show that electrostatic screening affects
both the time scale and spatial extent of confined particle motion,
whereas increasing swimming activity modifies both the oscillation
frequency and the accessible oscillatory region.

\begin{figure}[t]
    \centering
    \includegraphics[width=\textwidth]{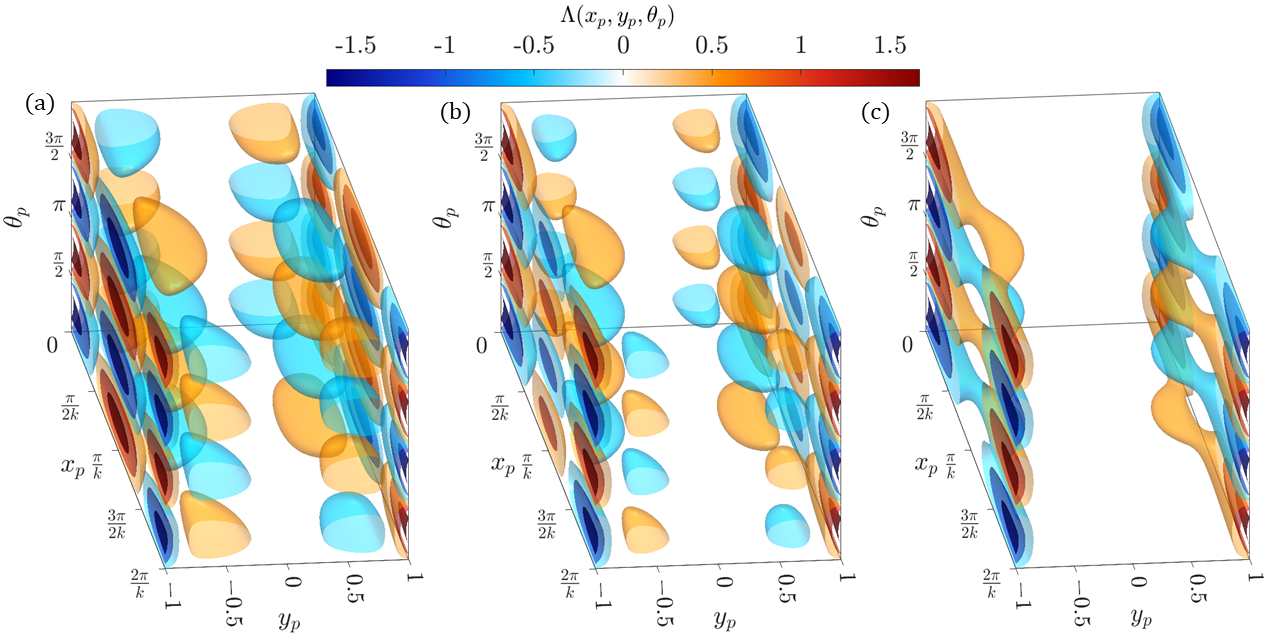}
    \caption{Isosurfaces of the phase-space volume expansion rate
    $\Lambda(x_p,y_p,\theta_p)$ for
    (a) $\eta=1$, (b) $\eta=0.8$, and (c) $\eta=0.6$.
    Warm and cool colours indicate local phase-space expansion
    ($\Lambda>0$) and contraction ($\Lambda<0$), respectively.
    As the uniform component of the zeta potential increases,
    the spatial organization of the expansion and contraction
    regions changes, with the interior structures becoming
    less prominent.}
    \label{fig:phase_volume}
\end{figure}

\begin{figure}[t]
    \centering
    \includegraphics[width=1\textwidth]{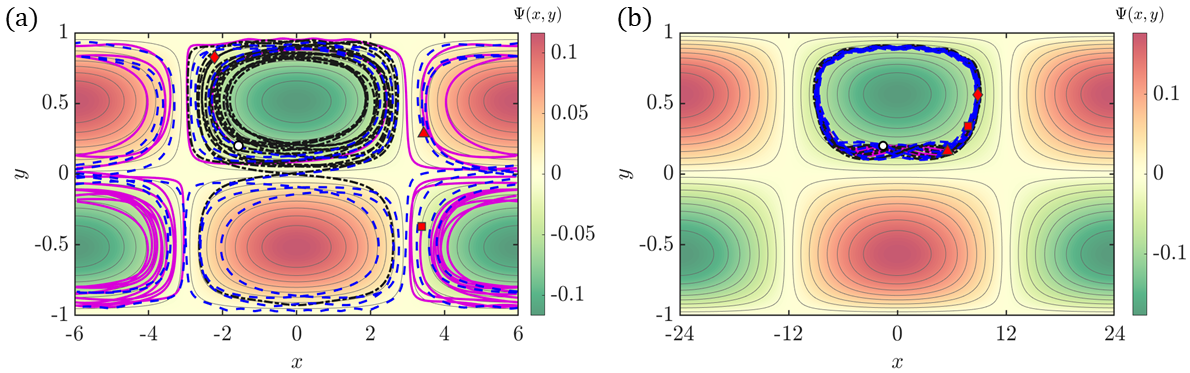}
    \caption{Particle trajectories superimposed on streamfunction contours $\Psi(x,y)$ for ($a$) $K = 10$, $k = \pi/6$ and ($b$) $K = 50$, $k = \pi/24$. The starting points are designated using (\whitemarker), while the end points are marked with a red filled square (\textcolor{red}{$\blacksquare$}) for $\alpha = 0.2$, a red filled diamond (\textcolor{red}{$\blacklozenge$}) for $\alpha = 0$, and a red filled upper triangle (\textcolor{red}{$\blacktriangle$}) for $\alpha = -0.2$. Trajectories are shown for $\alpha = 0.2$ (\textcolor{magenta}{magenta}), $\alpha = 0$ (black), and $\alpha = -0.2$ (\textcolor{blue}{blue}). The dimensionless swimming speed is fixed at $v_s = 0.01$. Altering the Debye parameter $K$ and wavenumber $k$ of the surface potential modifies the underlying vortex cellular structure, inducing a transition from localized circulatory dynamics to cross-stream spatial migration dependent on the particle shape parameter $\alpha$. See also Movies 1--6.}
  \label{fig:trajec_streamfun}
\end{figure}

\section{Spatially varying zeta potential}
\label{sec:patterned}


We now consider the purely patterned limit, $\eta=1$, for which the uniform component of the wall zeta potential vanishes and the
electroosmotic forcing is spatially periodic. In this limit, the
streamfunction in (\ref{eq:hydrosol_1}) reduces to
\begin{equation*}
\Psi(x,y)=f(y)\cos(kx),
\end{equation*}
where
\begin{equation*}
f(y)
=
B\sinh(ky)
+
Cy\cosh(ky)
-
\frac{\lambda\sinh(\lambda y)}
{K^2\cosh\lambda}.
\end{equation*}
Unlike the uniform zeta potential case, the resulting velocity field has both
streamwise and cross-stream components and varies periodically along the
channel. The spatial modulation therefore generates a cellular flow
consisting of counter-rotating vortices separated by fluid
separatrices (figure~\ref{fig:schematic}(b)).

This change in flow topology fundamentally alters the active particle
dynamics. In the uniform zeta potential limit, streamwise translational invariance
allows $x_p$ to decouple from the reduced dynamics in
$(y_p,\theta_p)$. For $\eta=1$, this invariance is broken because both
the background velocity and its gradients depend explicitly on $x_p$.
Consequently, the streamwise position $x_p$, cross-stream position $y_p$,
and orientation $\theta_p$ evolve as a fully coupled autonomous
three-dimensional system governed by
 (\ref{eq:dxdt_spheroid})--(\ref{eq:dthetadt_spheroid}). 

\subsection{Phase-space expansion and contraction}
\label{sec:phase_volume}

To examine how spatial patterning influences the organization of
active particle dynamics, we consider the local expansion and
contraction of the three-dimensional phase space
$(x_p,y_p,\theta_p)$. The local phase-space volume expansion rate,
$\Lambda(x_p,y_p,\theta_p)$, is defined as the divergence of
the dynamical vector field \citep{strogatz2024nonlinear},
\begin{equation}
\begin{aligned}
\Lambda(x_p,y_p,\theta_p)
&=
\frac{\partial\dot{x}_p}{\partial x_p}
+
\frac{\partial\dot{y}_p}{\partial y_p}
+
\frac{\partial\dot{\theta}_p}{\partial\theta_p}
\\
&=
-\alpha\left[
\left(
\frac{\partial^2\Psi}{\partial y_p^2}
-
\frac{\partial^2\Psi}{\partial x_p^2}
\right)\sin(2\theta_p)
+
2\frac{\partial^2\Psi}{\partial x_p\partial y_p}
\cos(2\theta_p)
\right].
\end{aligned}
\label{eq:volume_dynsys}
\end{equation}

Because the background electroosmotic flow is incompressible, the translational contributions to the phase-space divergence cancel, leaving
$\Lambda=\partial\dot{\theta}_p/\partial\theta_p$.
Thus, local phase-space expansion and contraction arise from the orientation-dependent response of non-spherical particles to the background velocity gradients. Positive values of $\Lambda$ indicate instantaneous expansion of infinitesimal
phase-space volumes, whereas negative values indicate
contraction. These local properties characterize the
deformation of neighbouring particle states in phase space but do not, by themselves, establish the existence of attractors or persistent particle trapping.

Figure~\ref{fig:phase_volume} illustrates how these regions evolve with the relative contribution of the patterned zeta potential. Although the present section focuses on the purely patterned limit ($\eta=1$), intermediate values $\eta=0.8$ and $0.6$ are included to illustrate the progressive influence of a uniform electroosmotic component. For $\eta=1$, the spatially varying velocity gradients produce alternating regions of phase-space expansion and contraction throughout the channel. As $\eta$ decreases, the increasing uniform component modifies the local shear and strain fields, progressively reorganizing these structures. In particular, the interior lobes become less
prominent, while pronounced expansion and contraction
regions persist near the walls. In the limiting case
$\eta=0$, streamwise translational invariance is recovered, and the dynamics reduces to the non-canonical Hamiltonian system discussed in \S\ref{sec:uniform}. Although local phase-space divergence need not vanish in this limit, the reduced dynamics preserves a weighted phase-space measure. Thus, varying $\eta$ connects the spatially structured dynamics of patterned electroosmotic flow to the integrable Hamiltonian dynamics of the uniform
zeta-potential limit.

Equation~(\ref{eq:volume_dynsys}) also reveals two important properties. First, $\Lambda$ vanishes identically for spherical particles ($\alpha=0$), demonstrating that local phase-space expansion and contraction arise from shape-dependent hydrodynamic reorientation. Second, integration over a complete orientation period gives
\begin{equation}
\int_0^{2\pi}
\Lambda(x_p,y_p,\theta_p)\,
\mathrm{d}\theta_p=0.
\label{eq:angular_avg_divergence}
\end{equation}
Hence, expansion and contraction balance under a uniform angular average, although this does not imply
phase-space volume preservation along individual
particle trajectories. The local divergence is also
independent of the swimming speed $v_s$; nevertheless,
self-propulsion influences the overall dynamics by
determining how particles move through these regions
of phase space.


\begin{figure}[t]
    \centering
    \begin{subfigure}[b]{1\linewidth}
        \centering
        \includegraphics[width=\linewidth, height=8cm]{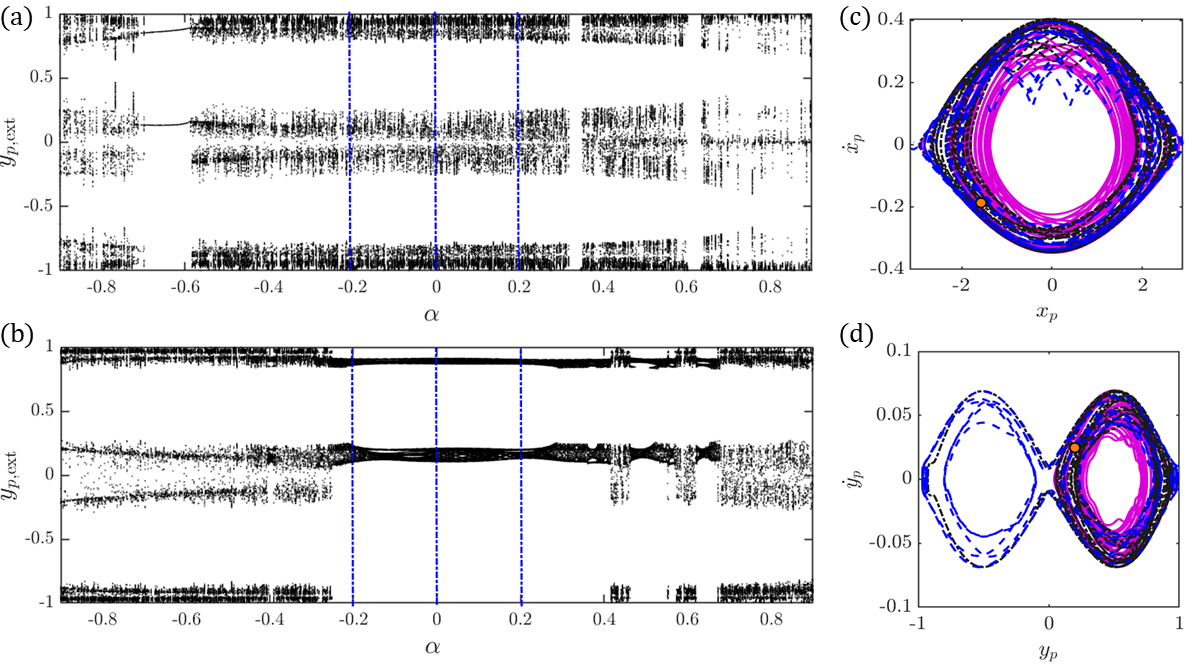}
        \label{fig:first}
    \end{subfigure}
    \vspace{0.5cm} 
    \begin{subfigure}[b]{1.0\linewidth}
        \centering
        \includegraphics[width=\linewidth, height=4cm]{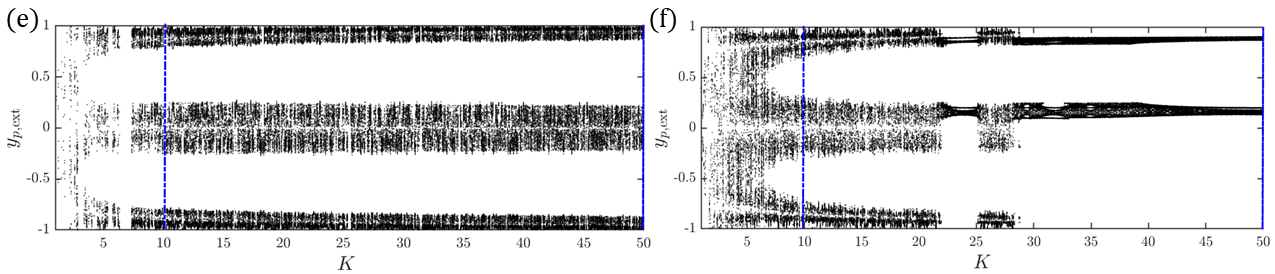}
        \label{fig:second}
    \end{subfigure}
\caption{Fixed-initial condition bifurcation diagrams and phase-space
projections for the purely patterned flow. Panels ($a,b$) show the
long-time cross-stream extrema $y_{p,\mathrm{ext}}$ versus $\alpha$ for
($a$) $K=10$, $k=\pi/6$ and ($b$) $K=50$, $k=\pi/24$.
Panels ($c,d$) show the corresponding $x_p$--$\dot{x}_p$ and $y_p$--$\dot{y}_p$ projections for $K=10$, $k=\pi/6$ and $\alpha=-0.8$ (\textcolor{magenta}{magenta}) ,$0$ (\textcolor{black}{black}) , $0.8$ (\textcolor{blue}{blue}). Panels ($e,f$) show $y_{p,\mathrm{ext}}$ versus $K$ for $\alpha = 0.2$, $k=\pi/6$, and $\pi/24$, respectively. Vertical dashed lines mark the parameter values corresponding to the representative trajectories
in figure~\ref{fig:trajec_streamfun}. For all panels, the initial condition is fixed to $(x_p(0),y_p(0),\theta_p(0))=(-\pi/2,0.2,0)$ and the swimming velocity is $v_s=0.01$.}
\label{fig:bifurcation}
\end{figure}

\subsection{Single-cell confinement and cross-cell motion}
\label{sec:patterned_bifurcation}

Having examined the local phase-space expansion and contraction,
we now consider how these dynamical features manifest in the
spatial trajectories of active particles. In particular, we
investigate whether the competition between self-propulsion,
vortex-induced advection, and flow-induced reorientation
confines particles within individual cellular regions or
allows them to migrate across neighbouring cells.

Figure~\ref{fig:trajec_streamfun} illustrates the particle dynamics
for two representative electroosmotic flow configurations and three
particle shapes, with trajectories superimposed on contours of the
background streamfunction. For $K=10$ and $k=\pi/6$
[figure~\ref{fig:trajec_streamfun}(a) and Movies~1--3],
the particles explore a relatively large portion of the cellular
flow, crossing fluid streamlines and, in some cases, moving between
neighbouring vortical regions. Unlike passive tracers, which remain
on individual streamlines in a steady two-dimensional flow, active
particles possess an additional swimming velocity that allows them
to move across the underlying recirculating structure. Whether this
motion leads to escape from a vortex depends on the particle
orientation, which evolves under the combined effects of local
vorticity and strain. Consequently, even at identical swimming
speeds, variations in the shape parameter $\alpha$ produce distinct
trajectories by modifying the response of the particle orientation
to the spatially varying velocity gradients.

For $K=50$ and $k=\pi/24$
[figure~\ref{fig:trajec_streamfun}(b) and Movies~4--6],
the particles instead remain predominantly localized within
individual vortical regions over the observation interval. In this
configuration, the swimming-induced displacement across fluid
streamlines is insufficient to produce sustained migration between
cells for the initial conditions considered, and the trajectories
largely follow the underlying recirculating flow. Particle shape
nevertheless affects the orbital paths through its influence on
flow-induced reorientation. Increasing $K$ corresponds to a thinner
electrical double layer relative to the channel height, while
decreasing $k$ increases the streamwise wavelength of the surface
pattern. Both changes modify the spatial distribution of the
electroosmotic velocity and its gradients, thereby altering the
balance between vortex circulation, swimming, and reorientation.


\begin{figure}[t]
	\centering
\includegraphics[width=1.03\textwidth]{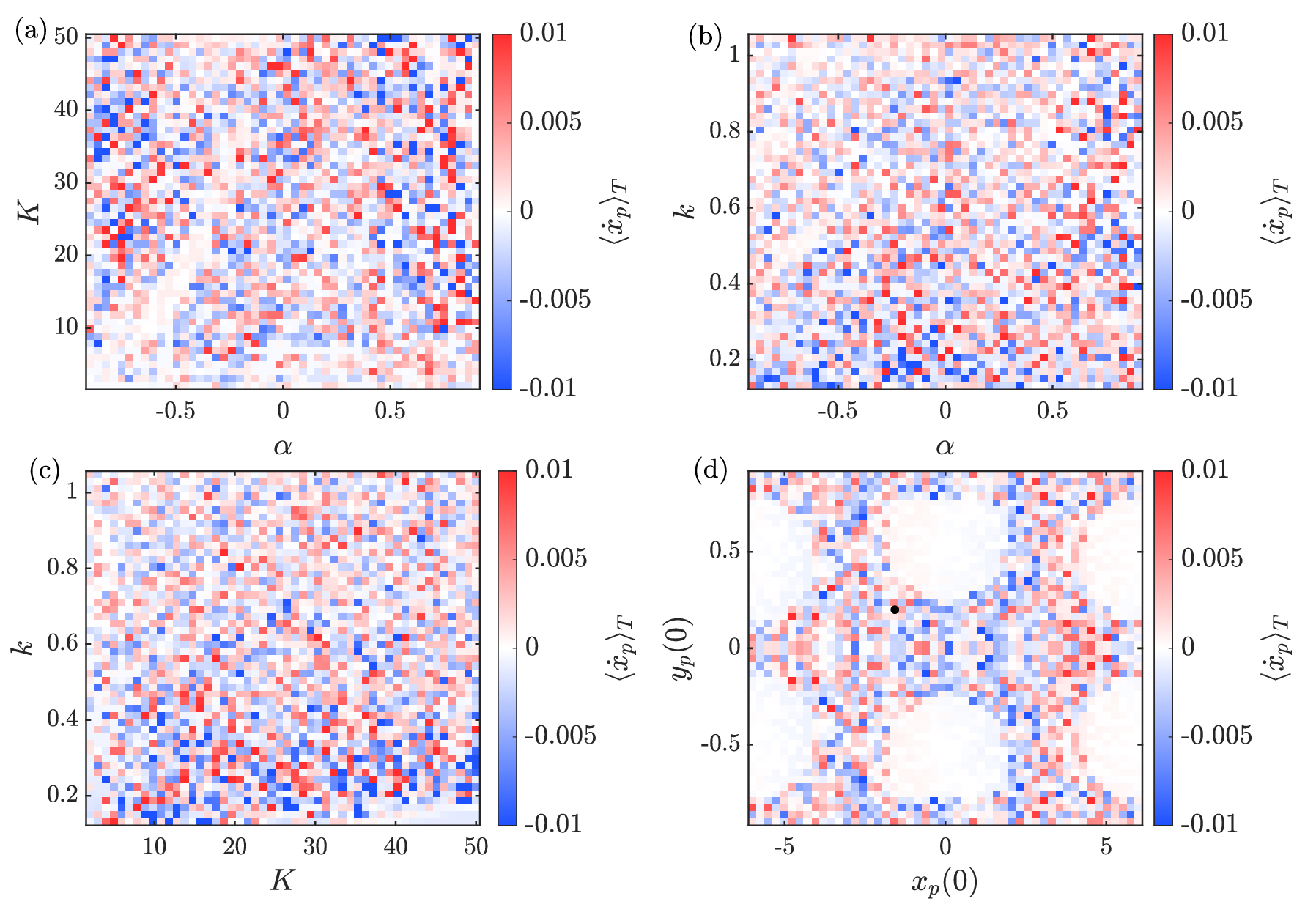}    

\caption{Time-averaged streamwise particle velocity
$\langle\dot{x}_p\rangle_T=[x_p(T)-x_p(0)]/T$
for the purely patterned electroosmotic flow ($\eta=1$).
Panels show the dependence on
(a) particle shape $\alpha$ and Debye parameter $K$ at $k=\pi/6$,
(b) $\alpha$ and surface-pattern wavenumber $k$ at $K=10$,
(c) $K$ and $k$ at $\alpha=0$, and
(d) initial particle position $(x_p(0),y_p(0))$ at
$\alpha=0$, $K=10$, and $k=\pi/6$.
For panels (a--c), the initial condition is fixed at
$(x_p(0),y_p(0),\theta_p(0))=(-\pi/2,0.2,0)$.
In panel (d), $\theta_p(0)=0$, and the black marker indicates
the reference initial position $(-\pi/2,0.2)$.
The swimming speed and integration time are fixed at
$v_s=0.01$ and $T=5000$, respectively.
Red and blue indicate positive and negative streamwise drift.}
\label{fig:drift_maps}

	\label{fig:activedriftmaps}
\end{figure}

To characterize these transport regimes beyond individual
trajectories, we construct bifurcation
diagrams by varying the particle shape parameter $\alpha$
and Debye parameter $K$, while keeping the remaining
parameters fixed in each sweep. For each parameter value,
the particle equations are integrated from
$(x_p(0),y_p(0),\theta_p(0))=(-\pi/2,0.2,0)$.
After discarding an initial transient, we record the
successive extrema of the cross-stream position,
denoted by $y_{p,\mathrm{ext}}$. These extrema provide
a compact representation of the cross-stream regions
visited by the particle. Discrete branches indicate
recurrent motion with a limited number of turning points,
whereas more broadly distributed extrema reflect
increasingly complex cross-stream dynamics.

Figures~\ref{fig:bifurcation}(a,b) show the variation
of $y_{p,\mathrm{ext}}$ with $\alpha$ for
$K=10$, $k=\pi/6$ and $K=50$, $k=\pi/24$,
respectively. Since the particle is initially located
in the upper half of the channel, extrema restricted
to positive $y_p$ indicate trajectories that remain
within the upper cellular region. In contrast, the
appearance of both positive and negative extrema
indicates cross-centreline motion, allowing the particle
to explore vortical regions on both sides of the channel.
The redistribution of these branches with $\alpha$
demonstrates how particle anisotropy modifies the
accessible cross-stream regions through its influence
on flow-induced reorientation.

The corresponding phase-space projections in
figures~\ref{fig:bifurcation}(c,d), obtained for
$K=10$, $k=\pi/6$ and $\alpha=-0.8,\,0,\,0.8$,
provide further insight into the underlying orbital
dynamics. The distinct structures in the
$x_p$--$\dot{x}_p$ and $y_p$--$\dot{y}_p$ planes
show that particle shape affects both the extent
and organization of recurrent motion. These
differences arise from the shape-dependent response
to the local strain field, which modifies the
swimming direction relative to the surrounding
vortex circulation.

Figures~\ref{fig:bifurcation}(e,f) examine the
influence of electrostatic screening by varying
$K$ at fixed $k=\pi/6$ and $k=\pi/24$,
respectively. For $k=\pi/6$, the extrema remain
broadly distributed over much of the parameter
range, indicating access to both upper and lower
cellular regions. In contrast, for $k=\pi/24$,
the diagrams exhibit more clearly separated
branches as $K$ increases, including regimes
where the extrema remain restricted to one
side of the channel. These changes reflect
the influence of electrostatic screening on
the spatial distribution of velocity gradients
and the resulting competition between
vortex-induced advection and particle
reorientation.

Taken together, these results demonstrate
that particle geometry and electrostatic
screening can reorganize the cross-stream
regions accessible to a swimmer with a
prescribed initial condition. The bifurcation
diagrams therefore complement the
representative trajectories by identifying
parameter regimes associated with
cross-centreline migration and spatial
localization.

The purely patterned electroosmotic flow has zero mean streamwise flux, yet self-propelled particles can exhibit a non-zero net displacement along the channel. 
Depending on the resulting trajectory, the particle may sample regions of opposing streamwise velocity unequally, leading to a finite net displacement even in the
absence of mean fluid advection. To quantify this behaviour, figure~\ref{fig:drift_maps} shows the finite-time mean streamwise velocity $\langle\dot{x}_p\rangle_T=[x_p(T)-x_p(0)]/T$ as a function of the governing parameters and initial particle position. Figures~\ref{fig:drift_maps}(a--c) examine the dependence on particle shape $\alpha$, Debye parameter $K$, and surface-pattern wavenumber $k$. The maps reveal interspersed regions of positive
and negative drift, indicating that both the direction and magnitude of streamwise transport are sensitive to these parameters. Particle anisotropy modifies the response to local strain, while electrostatic screening and pattern wavelength alter the spatial distribution of velocity gradients and vortex circulation. Their combined influence determines how
the particle moves through the cellular flow and accumulates streamwise displacement over the observation interval.

The sensitivity to the initial particle state is further illustrated in figure~\ref{fig:drift_maps}(d), where the initial streamwise and cross-stream positions are varied at fixed $\alpha=0$, $K=10$, and $k=\pi/6$. The resulting map exhibits distinct regions of weak, positive, and negative drift, demonstrating that particles with identical swimming properties can undergo different net displacements depending on their starting positions relative to the vortical structure. Together, these results show that spatially patterned electroosmotic flows can produce strongly parameter- and initial-condition-dependent active particle transport, despite having zero mean fluid flux. The observed sign changes suggest the possibility of controlling the direction of particle migration through electrostatic screening, surface-pattern wavelength, and particle geometry.

\begin{figure*}[t]
    \centering
    \includegraphics[width=\textwidth]{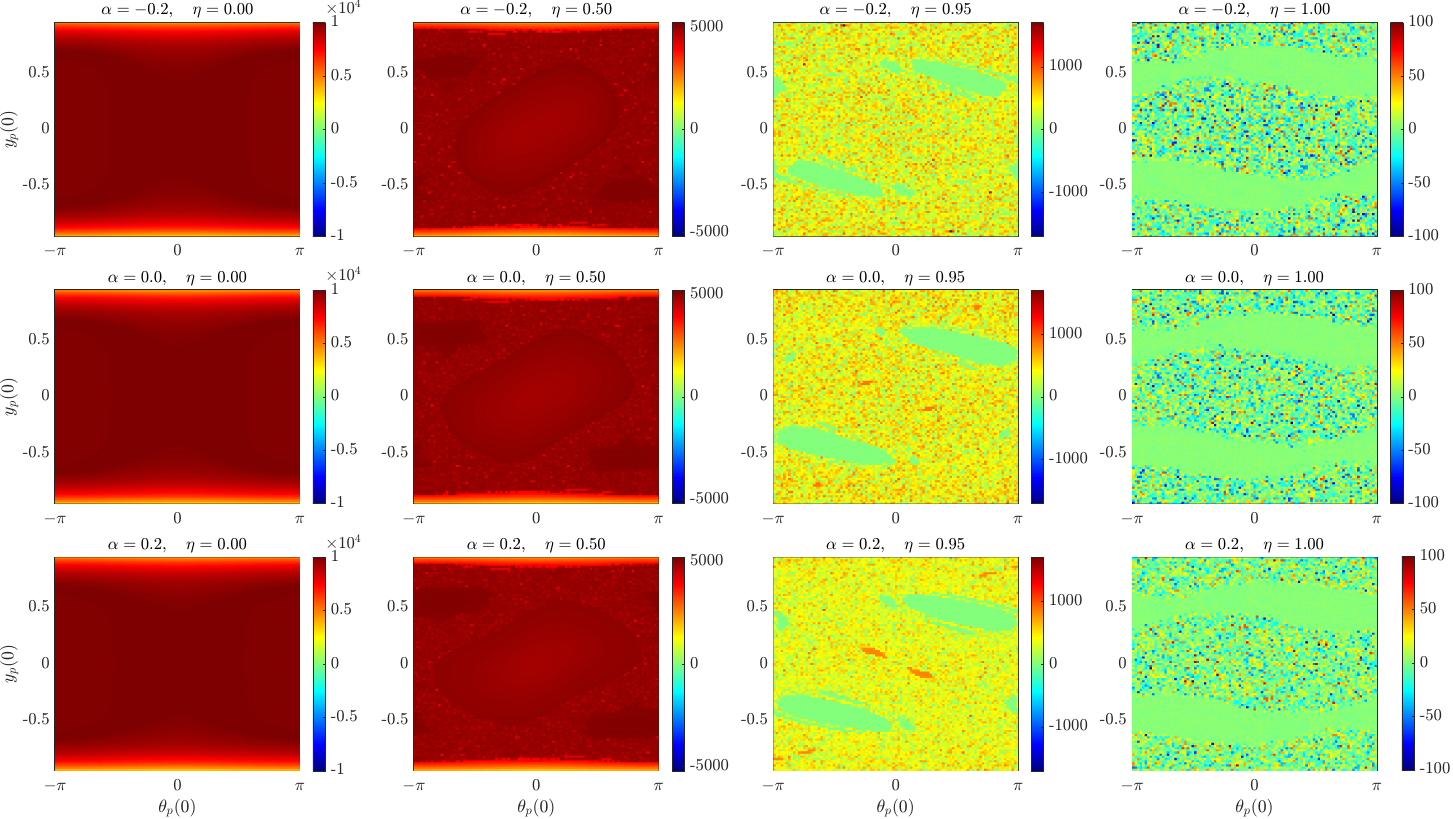}
    \caption{
    initial condition dependence of the streamwise particle displacement
    for $K=10$ and $k=\pi/6$. Each map shows the displacement
    $\Delta x_p=x_p(T)-x_p(0)$ as a function of the initial cross-stream
    position $y_p(0)$ and orientation $\theta_p(0)$. Rows correspond to
    particle shape parameters $\alpha=-0.2,\,0,\,0.2$, while columns
    correspond to $\eta=0,\,0.5,\,0.95,\,1$, spanning the transition
    from uniform to purely patterned zeta potential. The swimming speed,
    initial streamwise position, and integration time are fixed at
    $v_s=0.01$, $x_p(0)=-\pi/2$, and $T=10000$, respectively.
    The colour scale is chosen independently for each map to highlight
    variations in transport across the initial condition space.
    }
    \label{fig:IC_transport_K10}
\end{figure*}

\begin{figure*}[t]
    \centering
    \includegraphics[width=\textwidth]{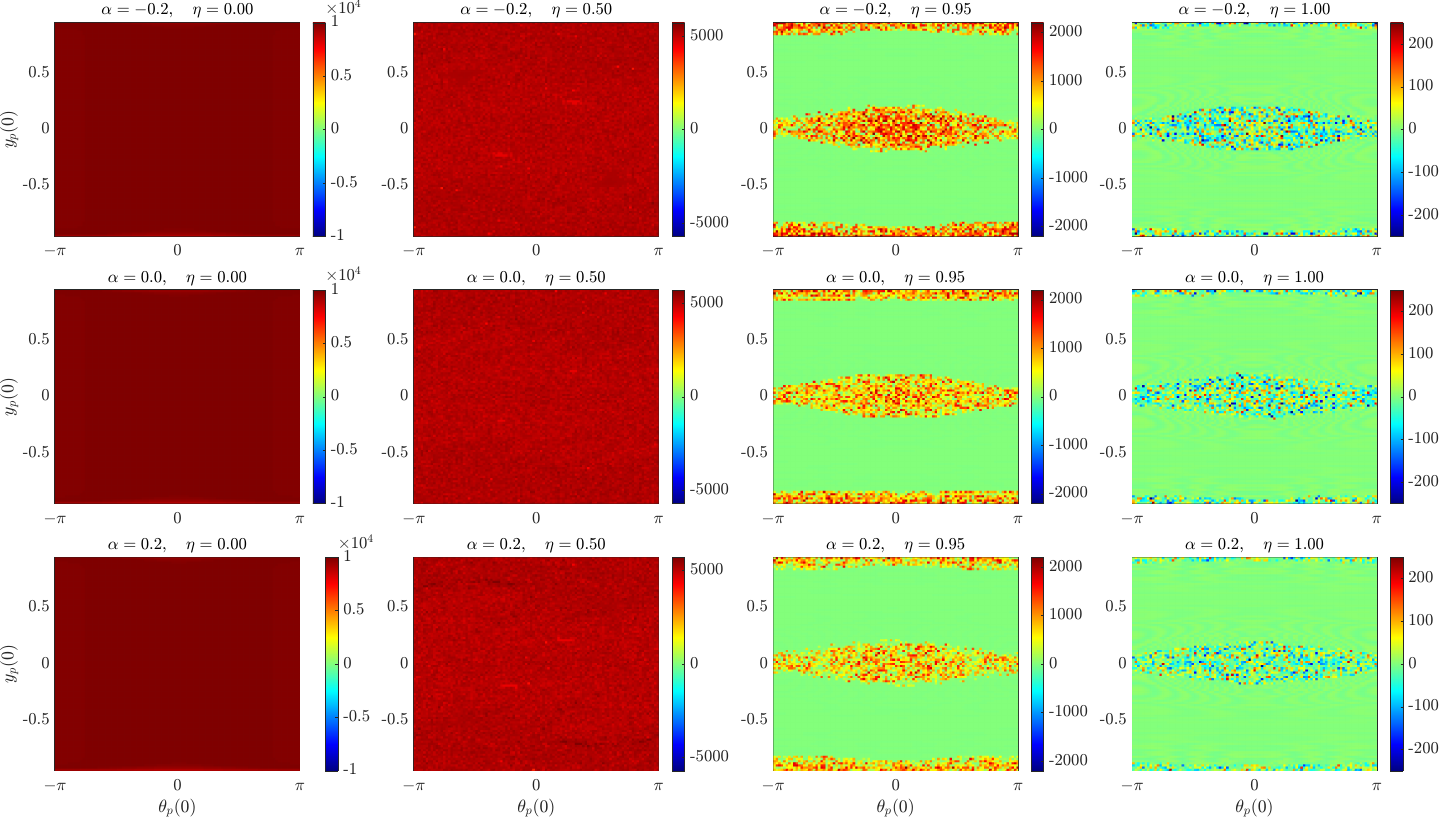}
    \caption{
    initial condition dependence of the streamwise particle displacement
    for $K=50$ and $k=\pi/24$. Each map shows the displacement
    $\Delta x_p=x_p(T)-x_p(0)$ as a function of the initial cross-stream
    position $y_p(0)$ and orientation $\theta_p(0)$. Rows correspond to
    particle shape parameters $\alpha=-0.2,\,0,\,0.2$, while columns
    correspond to $\eta=0,\,0.5,\,0.95,\,1$, spanning the transition
    from uniform to purely patterned zeta potential. The swimming speed,
    initial streamwise position, and integration time are fixed at
    $v_s=0.01$, $x_p(0)=-\pi/2$, and $T=10000$, respectively.
    The colour scale is chosen independently for each map to resolve
    variations in transport across the initial condition space.
    }
    \label{fig:IC_transport_K50}
\end{figure*}

\section{Combined uniform and spatially varying zeta potentials}
\label{sec:Combined}


We now consider the intermediate regime $0<\eta<1$, in which uniform and spatially varying components of the wall zeta potential coexist. The resulting electroosmotic flow combines mean streamwise advection with spatially varying recirculation, providing a continuous transition between the two limiting configurations examined in \S\ref{sec:uniform} and \S\ref{sec:patterned}.


To examine how this competition affects particle transport, we consider
an ensemble of particles initialized at the same streamwise position,
$x_p(0)$, while varying the initial cross-stream position
$y_p(0)$ and orientation $\theta_p(0)$. For each initial condition,
equations~(\ref{eq:dxdt_spheroid})--(\ref{eq:dthetadt_spheroid}) are
integrated to a fixed final time $T$, and we measure the net streamwise
displacement
\begin{equation}
\Delta x_p
=
x_p(T)-x_p(0).
\label{eq:net_displacement}
\end{equation}
Figures~\ref{fig:IC_transport_K10} and
\ref{fig:IC_transport_K50} show $\Delta x_p$ over the
$(\theta_p(0),y_p(0))$ initial condition plane for three representative
particle shapes and four values of $\eta$. The two figures correspond to
different combinations of the Debye parameter and surface-pattern
wavenumber, providing two distinct cellular-flow configurations with
which to examine the progression from uniform to patterned
electroosmosis.

For $\eta=0$, the displacement maps are comparatively smooth and show
only a weak dependence on the initial orientation over much of the
initial condition space. This behaviour is consistent with the
uniform-zeta dynamics of \S\ref{sec:uniform}, for which the streamwise
coordinate is dynamically decoupled from the reduced
$(y_p,\theta_p)$ system and the background electroosmotic flow provides
persistent downstream advection. Consequently, variations in
$\Delta x_p$ primarily reflect the cross-stream regions sampled by the
particle during its oscillatory motion.

As the patterned component is introduced, the transport becomes
increasingly sensitive to the initial particle state. At $\eta=0.5$,
substantial streamwise displacement persists, but the maps develop a
stronger dependence on both $y_p(0)$ and $\theta_p(0)$. This increased
heterogeneity reflects the competition between the uniform advective
component and the spatially varying cellular flow with particles initialized
at different positions and orientations sample different regions of the
vortical structure and therefore accumulate different streamwise
displacements.

This reorganization becomes more pronounced as $\eta$ approaches
unity. At $\eta=0.95$, extended regions of relatively small displacement
coexist with initial conditions that produce considerably larger
transport. The boundaries between these regions also depend on the
particle shape parameter $\alpha$, indicating that flow-induced
reorientation contributes to the partitioning of the initial condition
space. In the purely patterned limit, $\eta=1$, the uniform advective
component vanishes and the displacement is substantially reduced over
large portions of the initial condition plane. The remaining finite-time
displacement reflects motion within, and in some cases between, the
periodic vortical regions. 

Comparison of figures~\ref{fig:IC_transport_K10} and
\ref{fig:IC_transport_K50} shows that the detailed organization of the
transport regions also depends on the underlying electroosmotic flow
configuration. Although both cases exhibit the same broad progression
from predominantly advective transport at small $\eta$ to strongly
initial-condition-dependent motion as $\eta\rightarrow1$, the geometry
and extent of the low- and high-displacement regions differ substantially.

Overall, the intermediate regime provides a continuous dynamical
connection between the two limiting cases studied above. Increasing
$\eta$ progressively shifts the system from the Hamiltonian,
advection-dominated dynamics of the uniform-zeta limit towards the
vortex-mediated dynamics of the purely patterned flow. The relative
strength of the uniform and patterned surface potentials therefore acts
as an additional control parameter for tuning not only the magnitude of
particle transport, but also its sensitivity to particle shape and
initial conditions.

\section{Discussion and conclusion}
\label{sec:summary}


We have investigated the dynamics of self-propelled spheroidal particles in electroosmotic channel flows generated by uniform, spatially patterned, and combined zeta potential distributions. The uniform and patterned limits reveal qualitatively different mechanisms for organizing particle motion. For a uniform zeta potential, streamwise translational invariance reduces the dynamics to a two-dimensional system in cross-stream position and orientation. We showed that this system possesses a non-canonical Hamiltonian structure, with its phase space organized by an elliptic centre, a hyperbolic saddle, and the associated separatrix. In the spherical-particle limit, the system becomes canonical and admits a
nonlinear-pendulum interpretation. This description provides explicit expressions for both the small-amplitude oscillation frequency and the cross-stream
extent of the oscillatory dynamics. In particular, the oscillation frequency varies non-monotonically with the Debye parameter and reaches a maximum at an intermediate degree of electrostatic screening, demonstrating that the characteristic time scale of confined particle motion can be tuned through the electrical-double-layer properties.

Spatial patterning of the zeta potential fundamentally reorganizes the particle dynamics by generating a periodic array of counter-rotating electroosmotic vortices. The resulting loss of streamwise translational invariance couples the streamwise position, cross-stream position, and particle orientation into a fully three-dimensional dynamical system. Particle motion is then governed by the competition between vortex-induced advection, self-propulsion across fluid streamlines, and flow-induced
reorientation. Representative trajectories and bifurcation diagrams reveal distinct regimes of single-cell confinement and cross-cell migration, with particle shape and electrostatic screening controlling
the regions of the cellular flow accessible to the swimmer. Importantly, the absence of mean streamwise fluid transport does not necessarily imply zero net particle displacement. In the purely patterned limit, the electroosmotic flow has zero net flux, yet active particles can accumulate finite streamwise displacements through the interplay between swimming and spatially varying hydrodynamic reorientation. The computed drift maps reveal both positive and negative finite-time transport, with the direction and magnitude depending sensitively on particle shape $\alpha$, Debye parameter $K$, surface-pattern wavenumber $k$, and initial conditions. These observations suggest that active particles can sample regions of opposing streamwise flow unequally, producing an effective rectification of their motion over finite observation times. Thus, surface electrostatic patterning provides a potential means of controlling not only particle confinement and cross cell migration but also the direction and magnitude of streamwise transport.

The intermediate regime, $0<\eta<1$, provides a continuous connection
between these two limiting behaviours. As the patterned contribution is
increased, persistent streamwise advection associated with the uniform
component progressively competes with vortex-induced recirculation.
The resulting displacement maps show that transport becomes
increasingly sensitive to the initial cross-stream position and particle
orientation as $\eta$ approaches unity. Particle shape further modifies
the organization of these transport regions. Thus, varying the relative amplitudes of the uniform and patterned surface potentials does more than simply change the
magnitude of the electroosmotic flow, it reorganizes the particle
dynamics from predominantly advective transport towards strongly
initial-condition-dependent, vortex-mediated motion.

Taken together, these results provide a dynamical-systems perspective on
how surface electrostatics can reorganize active particle motion in
confined microflows. The uniform-zeta limit provides an analytically
tractable example of Hamiltonian dynamics, while spatial patterning
introduces cellular flow structures that enable localization and
cross-cell transport. Electrostatic screening, surface-pattern wavelength
and amplitude, swimming activity, and particle geometry therefore provide
complementary control parameters for modifying particle trajectories and
transport. More broadly, the ability to tune these dynamical regimes
through the wall zeta potential, without changing the channel geometry,
suggests a route towards electrokinetic control of active particle
confinement, migration, and transport in microfluidic environments.

\section*{Acknowledgements}
A.R. gratefully acknowledges Professor \href{https://www.me.iitb.ac.in/faculty/prof-salil-s-kulkarni}{Salil S. Kulkarni}, Indian Institute of Technology Bombay, India, for introducing him to the fundamental concepts and theoretical framework of nonlinear dynamics through the course ME 472, which provided a strong foundation for the present work. R.V. acknowledges the support of the Leverhulme Trust [Grant No. LIP-2020-014] and of the ERC Advanced Grant ActBio (funded as UKRI Frontier Research Grant EP/Y033981/1). A. P. gratefully acknowledges the support provided by the Department of Science and Technology, Government of India, through the project ANRF/ARG/2025/000835/ENS. 
\vspace{1mm}

\section*{Declaration of Interests}
The authors declare that there is no conflict of interest.


\FloatBarrier

\appendix
\section{Analytical solution for the electroosmotic flow using streamfunction approach}
\label{appenA}
To solve the momentum equations (\ref{eq:mom_dim}) for a purely periodic pressure field in $x$, we introduce a dimensionless stream function $\Psi(x,y)$ defined by $u = \partial \Psi/\partial y$ and $v = -\partial \Psi/\partial x$, which satisfies continuity (\ref{eq:cont_dim}) identically. Eliminating the pressure by taking the curl of the momentum equations yields the governing biharmonic equation,
\begin{equation}
    \nabla^4 \Psi = -(1-\eta)K^3 \frac{\sinh(K y)}{\cosh(K)}
    - \eta K^2 \frac{\lambda\sinh(\lambda y)}{\cosh(\lambda)}
    \cos(k x)
    \tag{A1}
    \label{eq:biharm_app}
\end{equation}
The flow is subjected to symmetry conditions at the centerline $y=0$ ($\partial^2 \Psi/\partial y^2 = 0$, $\partial \Psi/\partial x = 0$), no-slip and no-penetration conditions at the upper wall $y=1$ ($\partial \Psi/\partial y = 0$, $\partial \Psi/\partial x = 0$), and periodicity in $x$ with period $2\pi/k$. Seeking a solution compatible with these boundary conditions and the spatial forcing structure, we adopt the separable ansatz $\Psi(x,y) = \Psi_0(y) + g(y) \cos(k x)$, which reduces (\ref{eq:biharm_app}) to the ordinary differential equation
\begin{subequations}
    \begin{align}
    \frac{d^4 \Psi_0}{dy^4} = -(1-\eta) K^3 \frac{\sinh(Ky)}{\cosh(K)}
    \tag{A2(a)}\label{eq:A2a}\\
    \left( \frac{d^2}{dy^2} - k^2 \right)^2 g(y) = - \eta K^2 \frac{\lambda\sinh(\lambda y)}{\cosh(\lambda)}
    \tag{A2(b)}\label{eq:A2b}
    \end{align}
\end{subequations}
Integrating (\ref{eq:A2a}) four times and applying the boundary conditions yields the classical plug-flow stream function:
\begin{equation}
    \Psi_0(y) = (1-\eta) \left[ y - \frac{\sinh(K y)}{K \cosh(K)}\right]
    \tag{A3}
    \label{eq:psisol0}
\end{equation}
Similarly, subject to symmetry and no-slip boundary conditions we also solve (\ref{eq:A2b}) decomposing it into homogeneous and particular solutions, such that the homogeneous solution satisfying the odd-symmetry conditions at the centerline. Additionally, without loss of generality let us assume the particular solution takes the form $g_p(y) = A \sinh(\lambda y)$. Ultimately combining (\ref{eq:psisol0}) yields the closed form solution of the streamfunction yields:
\begin{equation}
    \Psi(x,y) = (1-\eta)\left[y - \frac{\sinh(Ky)}{K \cosh(K)}\right] + \eta \left[ B \sinh(ky) + C y \cosh(ky) - \frac{\lambda\sinh(\lambda y)}{K^2\,\cosh(\lambda)} \right] \cos(kx)
    \tag{A4}
    \label{eq:psisol}
\end{equation}
Henceforth, the constants $B$ and $C$ in (\ref{eq:psisol}), are given in terms of the known parameters which are evaluated by solving a linear system subjected to the boundary conditions, as prescribed below:
\begin{equation}
    B = \frac{\lambda\tanh(\lambda) \left(\cosh(k) + k \sinh(k)\right) - \lambda^2 \cosh(k)}{K^2 \left(\sinh(k) \cosh(k) - k\right)}, \quad
    C = \frac{\lambda^2 \sinh(k) - k \cosh(k) \lambda\tanh(\lambda)}{K^2 \left(\sinh(k) \cosh(k) - k\right)}
    \tag{A5}
\end{equation}
\section{Derivation of the non-canonical Hamiltonian structure}
\label{appenB}
To establish the Hamiltonian formulation for the reduced two-dimensional dynamical system ($\eta = 0$) presented in (\ref{eq:dy_1}--\ref{eq:dtheta_1}), we seek an invariant phase-space density function $\mu(\theta_p)$ that renders the system divergence-free. The continuity condition for the weighted vector field $( \mu f, \mu g )$ is given by
\begin{equation}
    \frac{\partial}{\partial y_p} (\mu f) + \frac{\partial}{\partial \theta_p} (\mu g) = 0, \quad \left[f(y_p, \theta_p) = v_s \sin\theta_p, \, g(y_p, \theta_p) = -\frac{K}{2} \frac{\sinh(K y_p)}{\cosh(K)} [\alpha \cos(2\theta_p) - 1] \right]
    \label{eq:density}
    \tag{B1}
\end{equation}
Since $f(y_p, \theta_p)$ is independent of $y_p$, substituting the components into (\ref{eq:density}) yields the reduced ordinary differential equation for $\mu(\theta_p)$:
\begin{equation}
    \frac{\mathrm{d}}{\mathrm{d}\theta_p} \Big( \mu(\theta_p) \big[\alpha \cos(2\theta_p) - 1\big] \Big) = 0 \implies \mu(\theta_p) = \frac{1}{1 - \alpha \cos(2\theta_p)}
    \label{eq:density_func}
    \tag{B2}
\end{equation}
Equipped with the density factor $\mu(\theta_p)$ from (\ref{eq:density_func}), the underlying non-canonical Hamilton equations are defined as: 
\begin{subequations}
\label{eq:B3}
\begin{align}
\frac{1}{\mu(\theta_p)}
\frac{\partial H}{\partial \theta_p}
&= v_s \sin\theta_p \label{eq:B3a} \tag{B3(a)}\\
-\frac{1}{\mu(\theta_p)}
\frac{\partial H}{\partial y_p}
&= -\frac{K}{2}
\frac{\sinh(K\,y_p)}{\cosh(K)}
\big[\alpha \cos(2\theta_p)-1\big] \label{eq:B3b} \tag{B3(b)}
\end{align}
\end{subequations}
Integrating (\ref{eq:B3a}--\ref{eq:B3b}) with respect to $y_p$ and $\theta_p$ yields the separable Hamiltonian
\begin{equation}
    H(y_p, \theta_p) = -\frac{\cosh(K y_p)}{2 \cosh(K)} + \int \frac{v_s \sin\theta_p}{1 - \alpha \cos(2\theta_p)} \, \mathrm{d}\theta_p
    \label{eq:hamiltonian_general}
    \tag{B4}
\end{equation}
The evaluation of the angular integral in (\ref{eq:hamiltonian_general}) depends explicitly on the anisotropy parameter $\alpha \in (-1, 1)$. Performing the piecewise anti-derivation across all physical regimes of $\alpha$ and $v_s \in [0, \infty)$ produces
\begin{equation}
    \int \frac{v_s \sin\theta_p}{1 - \alpha \cos(2\theta_p)} \, \mathrm{d}\theta_p = \begin{cases} 
        \displaystyle -\frac{v_s}{\sqrt{-2\alpha(1+\alpha)}} \arctan\left( \sqrt{\frac{-2\alpha}{1+\alpha}} \cos\theta_p \right), & -1 < \alpha < 0 \\[12pt]
        \displaystyle -v_s \cos\theta_p, & \alpha = 0 \\[12pt]
        \displaystyle -\frac{v_s}{\sqrt{2\alpha(1+\alpha)}} \mathrm{artanh}\left( \sqrt{\frac{2\alpha}{1+\alpha}} \cos\theta_p \right), & 0 < \alpha < 1 \\[12pt]
    \end{cases}
    \label{eq:potential_piecewise}
    \tag{B5}
\end{equation}

\bibliographystyle{jfm2}
\begingroup
\def\bibfont{\small}
\setlength{\bibsep}{1.1ex}
\bibliography{activecite}
\endgroup	
\end{document}